\documentclass{article}

{}
{}
{}

\usepackage{amsmath} 
\usepackage{amsthm} 
\usepackage{cite} 
\usepackage[numbers,compress,sort]{natbib}
\usepackage{caption}
\usepackage{url}
\usepackage{color}
\usepackage{longtable}
\usepackage{multirow}
\usepackage{hyperref} 
\usepackage{booktabs}
\usepackage[pdftex]{graphicx}
\usepackage{graphicx}
\usepackage{subfigure}
\usepackage{algorithm}
\usepackage{algorithmic} 
\usepackage{hyperref}
\usepackage{authblk}
\usepackage[margin=1.2cm]{geometry}
\usepackage{multicol}
\usepackage{blindtext}
\usepackage{epstopdf}
\usepackage{float}
\usepackage{varwidth}

\begin{document}

\title{A Combination Model for Time Series Prediction using LSTM via Extracting Dynamic Features Based on Spatial Smoothing and Sequential General Variational Mode Decomposition}
\author[1]{Jianyu Liu}
\author[2\thanks{This work was supported by the Natural Science Foundation of Jiangxi Province (No. 20232ACB202005).}]{Wei Chen}

\author[3]{Yong Zhang}
\author[4]{Zhenfeng Chen}
\author[5]{Bin Wan}
\author[6]{Jinwei Hu}

\affil[1]{{ JiangXi University of Finances and Economics} \newline  *Email: chenwei@jxufe.edu.cn }
\affil[1]{{ JiangXi University of Finances and Economics} \newline  }
\affil[1]{{ JiangXi University of Finances and Economics} \newline  }
\affil[1]{{ JiangXi University of Finances and Economics} \newline  }
\affil[1]{{ JiangXi University of Finances and Economics} \newline  }
\affil[1]{{ JiangXi University of Finances and Economics} \newline  }

\setcounter{Maxaffil}{0}
\renewcommand\Affilfont{\itshape\small}

\date{}

\maketitle

\begin{abstract}
In order to solve the problems such as difficult to extract effective features and low accuracy of sales volume prediction caused by complex relationships such as market sales volume in time series prediction, we proposed a time series prediction method of market sales volume based on Sequential General VMD and spatial smoothing Long short-term memory neural network (SS-LSTM) combination model. Firstly, the spatial smoothing algorithm is used to decompose and calculate the sample data of related industry sectors affected by the linkage effect of market sectors, extracting modal features containing information via Sequential General VMD on overall market and specific price trends; Then, according to the background of different Market data sets, LSTM network is used to model and predict the price of fundamental data and modal characteristics. The experimental results of data prediction with seasonal and periodic trends show that this method can achieve higher price prediction accuracy and more accurate accuracy in specific market contexts compared to traditional prediction methods Describe the changes in market sales volume.
\end{abstract}

\begin{multicols}{2}

\section{Introduction}

Due to the importance of the market, historical sales volume data and various parameter indicators of each product are analyzed and mined\cite{1}, and relevant theories and algorithm models are used to predict
Measuring the trend of the market has important theoretical significance and social value\cite{2}. As a complex and time-varying huge system\cite{3}, the price fluctuation of market sales often shows strong non-linearity\cite{4}. Therefore, it is difficult to obtain valuable information from massive Market data to serve decision-making. In view of the numerous Market data indicators and the existence of high-dimensional nonlinear characteristics\cite{5}, the current mainstream market sales forecasting methods mainly combine data features extraction and model prediction process to study\cite{6}.

By reason of the numerous and high-dimensional nonlinear characteristics of market sales data indicators, current methods for predicting sales mainly combine data features extraction with model prediction processes for research\cite{7}. For example, N. E. Huang et al. adopt empirical mode decomposition algorithm (EMD) to decompose the time series, then they can successfully Prediction market price trend based on the modal information obtained from the decomposition\cite{8}, which is an operation based on the oscillation characteristics of signal contour to extract components from the time domain\cite{9}. Konstantin Dragomiretskiy from UCLA proposed the Variational Mode Decomposition (VMD) method\cite{10}, which estimates various signal components by solving frequency domain variational optimization problems. Dynamic mode decomposition was first used to analyze the dynamic processes of fluids (such as water flow)\cite{11}. It can decompose complex flow processes into low rank spatiotemporal features. Dynamic mode decomposition can also be used to analyze complicated time series in general. K. Karl Pearson's principal component analysis (PCA) for non random variables is a statistical method that transforms a set of potentially correlated variables into a set of linearly uncorrelated variables through orthogonal transformations\cite{12}. This can be used for features extraction and data dimensionality reduction, combined with some commonly used neural networks for modeling research on time series. Most of the current time series forecasting methods directly select Market data as the research object, without considering the influence and potential related factors of the overall trend of related industry plates under the plate linkage effect; In terms of features extraction, operations such as dimensionality reduction and transformation of data lack business interpretability; The prediction model usually selects time series analysis methods based on statistical theory and traditional machine learning methods, which have shortcomings in the accuracy of nonlinear, non-stationary, seasonal, periodic, and trendy time series prediction and the utilization of time series information.

However, currently, the methods dedicated to pattern decomposition problems are very specific, and most are designed to solve certain types of problems with strong and unique limitations, making it difficult to extend to other fields.

In the field of signal processing, non-stationary signal separation is an important topic. Due to the loss of stability, ordinary tools cannot solve this problem. Several methods have been proposed and some progress has been made, mainly focusing on the frequency domain and rarely on the time domain. The most common methods are EMD, VMD, DMD, e.t.c.

Empirical Mode Decomposition (EMD) is a signal processing technique proposed by Huang in 1998. Based on the local characteristics and adaptability of the signal, the algorithm decomposes the non-stationary signal into a set of intrinsic mode functions (IMF), each IMF represents a local feature of the signal. The core idea of EMD is to iteratively extract local extreme points in the data and construct upper and lower envelopes to obtain IMF until the conditions that meet specific stopping criteria are met. However, the EMD algorithm has some limitations, such as modal overlap, modal mixing, and convergence, which limit its effectiveness and stability in practical applications\cite{13}.

Variational Mode Decomposition (VMD) is a signal processing technique proposed by Konstantin Dragomiretskiy in 2014. Based on the Variational principle, the algorithm decomposes the non-stationary signal into a group of modal functions with different frequency and amplitude modulation. The idea of VMD is to decompose signals by finding a combination of modal functions with the minimum Total variation, where each modal function is part of the signal and is locally modulated in frequency and amplitude. However, the VMD algorithm may face difficulties in dealing with high noise and limited data length, and is also sensitive to the selection of different signal types and parameters. These limitations need to be considered in practical applications\cite{14}.

In this paper, we used a new algorithm named Sequential General Variational Mode Decomposition Method(SGVMD) proposed by Chen\cite{SGVMD}. This method can fetch one component from one decomposition, without prior knowledge or estimation of the number of intrinsic modes. This function is achieved by modifying the optimization form based on VMD. This advantage can widen the application field of the VMD method in non-stationary signal decomposition. This new algorithm performs as a series of adaptive filters to extract principal components from the mixture one by one. To reduce the end effect, an approach called Principal Component Restoring(PCR) is introduced\cite{16}. We extract the trend line and several principal components from the end regions, then conduct an elongation  operation based on such information for the signal. In the proposed method, we would do decomposition for
the elongated signal, instead of the mirrored one in VMD. Also, the extension of its application to two-dimensional
signal decomposition can be realized with some essential
modifications.

Dynamic Mode Decomposition (DMD) is a signal processing technique used for analyzing dynamic systems\cite{17}, first proposed by Peter Schmid in 2010. This algorithm can extract the dynamic modes of the system from measurement data and use these modes to predict and analyze system behavior. The idea of DMD is to represent the evolution process of the system as a set of eigenvectors and corresponding eigenvalues, which can be obtained by calculating the Linear map between data snapshots. However, DMD algorithm is highly sensitive to high-dimensional data and noise, and has limited adaptability to nonlinear systems\cite{wu}. These limitations need to be considered in practical applications.

But if turn our attention back to the field of array signal processing in the 20th century, we will find that such so-called "dynamic" ideas have already been widely applied. The development of array signal processing algorithms has gone through multiple important milestones. The Multiple Signal Classification algorithm (MUSIC) was proposed by Schmidt in 1979. It uses the eigen Vector space of the signals received by the array to perform subspace decomposition\cite{18}, so as to achieve high-resolution direction estimation of the signal source. Estimation of Signal Parameters via Rotational Invariance Techniques (ESPRIT) algorithm was proposed by Roy in 1986\cite{19}. It uses the Rotational invariance of the signal received by the array to estimate the direction of the signal source through Matrix decomposition. Spatial smoothing algorithm is a commonly used technique in array signal processing, used to improve the accuracy and stability of direction estimation by smoothing signal subspaces to suppress the influence of noise and uncorrelated signals. However, these algorithms may face performance degradation issues when facing challenges such as low signal-to-noise ratio, coherent signals, and imperfect array structures. Further improvement and optimization are still needed for complex environments and real-time applications.

The ESPRIT and spatial smoothing MUSIC algorithms in the field of array signal processing share similar ideas with the so-called dynamic mode decomposition. In array signal processing, if the received signal is reflected back onto the receiving antenna array by a certain obstacle, the signal will generate a certain amplitude attenuation and phase shift, and such a signal is called a coherent signal. If the wireless channel or reception environment is very harsh, the received signals may have strong correlation or even complete coherence. However, such completely coherent signals can later be considered as complete dictionaries in the sparse signal processing that emerged in the early 21st century, that is, a large number of unrelated bases, or we can become signals with sparse properties composed of frames. But we can still capture the features or even so-called "dynamic features" we need from such highly correlated signals. In response to the issues identified in the above research, we can apply such ideas to feature extraction and prediction of time series.

Therefore, we proposed a time series prediction method of market sales volume based on Sequential General VMD and spatial smoothing Long short-term memory neural network (SS-LSTM) combination model. Starting from the linkage effect of related data sectors, this article can search for dynamic features. Firstly, different algorithms are used to construct datasets, and spatial smoothing algorithms are used to decompose and calculate the sales sample data under the related industry sectors. Modal features representing the overall market trend of the industry sector and the price trend information of sales themselves are extracted by SGVMD. Subsequently, the LSTM neural network was used to have a good fitting and forecasting ability for nonlinear time series, and LSTM models were constructed to learn sales fundamental data and modal characteristics under different market backgrounds. Finally, the sales forecasting method and its fusion model based on the spatial smoothing and sequential general variational mode decomposition using Long short-term memory network (SS-LSTM) model were proposed.

\section{Methodology}
\subsection{Space Smoothing Algorithm}

Spatial smoothing in array signal processing is a technique used to increase the correlation of sampled data in a multidimensional dataset by adding windows sampling to appropriate dimensions for analysis and to reduce interference from coherent signals in the analysis. This is a method of identifying patterns in data and expressing them to highlight similarities (patterns) and differences (weights of patterns) in the data. The principal component decomposition of data is the most effective choice in linear decomposition, as it retains the highest possible kinetic energy on average and extracts significant relevant feature representations in low dimensional space.

Suppose there is an antenna array with n elements for receiving signals, and the transmitting source has m elements. Generally speaking, m is not greater than n. The so-called direction of arrival (DOA) is the angle between the incoming signal and the antenna unit. In array signal processing, people usually calculate the DOA. However, such a uniform linear array receiving model is likely to be interfered by some obstacles, resulting in coherent signals with amplitude attenuation and phase offset, although at the same frequency. According to the theory of subspaces, such a large antenna array can be divided into sub arrays composed of L antenna elements in each group. So the schematic diagram of this antenna array model is

\begin{figure}[H]
	\centering	
	\includegraphics[width=\linewidth]{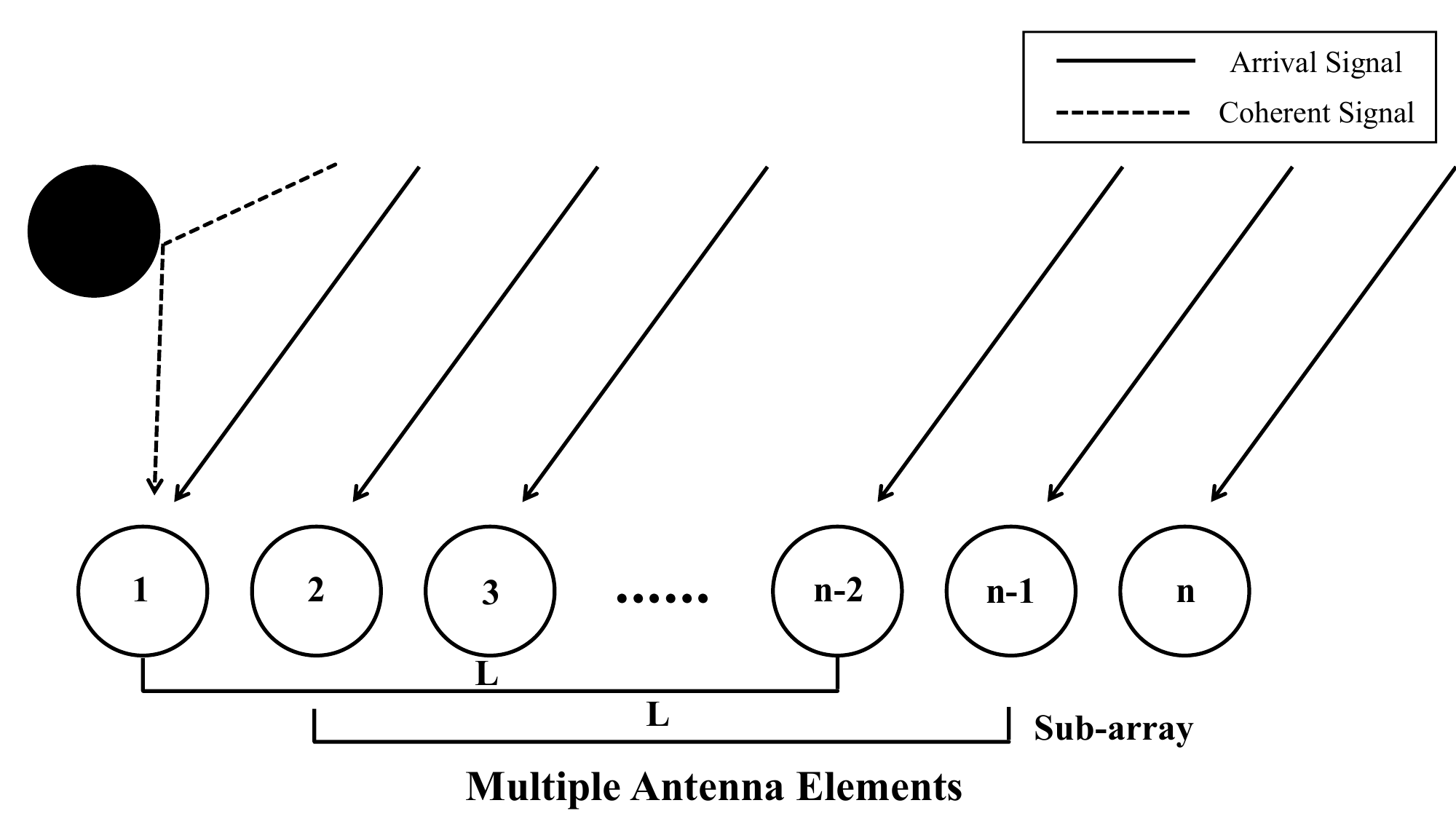}
	\caption{Antenna array receives arrival signals and its coherent signals.}
	\label{1}
\end{figure}

By taking the sum of the signals received by each antenna array as one row of the matrix, it is natural that the received signals will have n rows. Assuming the distance between each antenna element in a uniform linear array is $d$, the first element receives an incoming signal with no phase shift version, and the subsequent elements receive a phase shifted version of the incoming signal.
\begin{equation} 
\mathnormal{X}(t-\Delta)=\mathnormal{X}(t)e^{-2\pi(d/\lambda_0)sin\theta}=\mathnormal{X}(t)e^{j\phi}
\end{equation}
Where time delay $\Delta=d\sin\theta/C$ and velocity of light $C=f_0\lambda_0$. And simplify the distance between each element of the uniform linear array and the information of direction of arrival into $\phi$. In this way, the time delay of the wave signal under the narrowband assumption is converted into phase shift. For each antenna array element, assuming there are $m$ arrival signals and the snapshot is taken $n$ times, it will form the $n\times m$ data matrix $S$ of the arrival signal.
\begin{equation} 
	\begin{pmatrix}
	\mathnormal{X_1} \\
	\mathnormal{X_2} \\
	\vdots\\
	\mathnormal{X_n} \\
	\end{pmatrix} = 
	\begin{pmatrix}
		1 & 1 & \cdots & 1 \\
		e^{j\phi_1} & e^{j\phi_2} & \cdots & e^{j\phi_m} \\
		e^{j2\phi_1} & e^{j2\phi_2} & \cdots & e^{j2\phi_m} \\
		\vdots & \vdots & \ddots & \vdots \\
		e^{j(n-1)\phi_1} & e^{j(n-1)\phi_2} & \cdots & e^{j(n-1)\phi_m} 
	\end{pmatrix}
	\begin{pmatrix}
		\mathnormal{S_1} \\
		\mathnormal{S_2} \\
		\vdots\\
		\mathnormal{S_m} \\
	\end{pmatrix} +
	\begin{pmatrix}
		\mathnormal{N_1} \\
		\mathnormal{N_2} \\
		\vdots\\
		\mathnormal{N_m} \\
	\end{pmatrix}
\end{equation}
Assuming the discussion is about a uniform linear array, as shown in Fig. \ref{1}. If each sub-array arranged has $L$ elements, many windows can be smoothed out.
\begin{equation} 
	\begin{split}
	\mathnormal{X_1} &= A_1(\theta)S + \mathcal{N}_1\\
	\mathnormal{X_2} &= A_2(\theta)S + \mathcal{N}_2\\
	&\cdots    \quad\cdots \\
	\mathnormal{X_{n-L+1}} &= A_{n-L+1}(\theta)S + \mathcal{N}_{n-L+1}\\
	\end{split}
\end{equation}
Where $A_i(\theta)$ carries information about the direction of arrival of the signal, and each $A_i(\theta)$ changes. This type of $A_i(\theta)$ with a certain "ridge" shape and high correlation with changes in $i$ will be used multiple times in this article.
\begin{equation} 
	\begin{split}
	A_1(\theta)&=
	\begin{pmatrix}
		1 & 1 & \cdots & 1 \\
	e^{j\phi_1} & e^{j\phi_2} & \cdots & e^{j\phi_m} \\
	\vdots & \vdots & \ddots & \vdots \\
	e^{j(L-1)\phi_1} & e^{j(L-1)\phi_2} & \cdots & e^{j(L-1)\phi_m} 
	\end{pmatrix}\\
	A_2(\theta)&=
\begin{pmatrix}
	\quad e^{j\phi_1} & \quad e^{j\phi_2} & \quad \cdots & \quad e^{j\phi_m} \\
	\quad e^{j2\phi_1} & \quad e^{j2\phi_2} & \quad \cdots & \quad e^{j2\phi_m} \\
	\quad \vdots & \quad \vdots &  \quad \ddots & \vdots \\
	\quad e^{jL\phi_1} & \quad e^{jL\phi_2} & \quad \cdots & \quad e^{jL\phi_m} 
\end{pmatrix}\\
	&\quad \quad \quad \quad \quad \quad \quad \cdots    \quad\cdots \\
	A_{n-L+1}(\theta)&=
\begin{pmatrix}
	1 & 1 & \cdots & 1 \\
	e^{j\phi_1} & e^{j\phi_2} & \cdots & e^{j\phi_m} \\
	\vdots & \vdots & \ddots & \vdots \\
	e^{j(n-1)\phi_1} & e^{j(n-1)\phi_2} & \cdots & e^{j(n-1)\phi_m} 
\end{pmatrix}\\
	\end{split}
\end{equation}
It can be observed that there are certain relationships between adjacent A, but it cannot be ignored that this is only under the condition of a uniform linear array.
\begin{equation}
	\begin{split}
	A_2(\theta) &= A_1(\theta)B\\
	A_3(\theta) = A_2&(\theta)B = A_1(\theta)B^2\\
	\cdots    &\quad\cdots \\
	A_k(\theta) &= A_1(\theta)B^{k-1}
	\end{split}
\end{equation}
Where $B$ is a diagonal array, and for a known antenna array distance, a definite $B$ can be obtained:
\begin{equation}
	B = diag(e^{j\phi_1},\cdots,e^{j\phi_m})
\end{equation}
After simply processing the received arrival signal, create a correlation matrix for the k-th sub-array:

\begin{equation}
	\begin{split}
	\mathnormal{R_\mathnormal{X}^{(k)}}&=\mathnormal{E}(\mathnormal{X}_k\mathnormal{X}_k^H)\\
	&=A_k(\theta)\mathnormal{R_\mathnormal{S}}A_k^H(\theta)+\sigma^2\mathnormal{I}\\
	&=A_1(\theta)B^{k-1}\mathnormal{R_\mathnormal{S}}(B^{k-1})^{H}A_1^H(\theta)+\sigma^2\mathnormal{I}	
	\end{split}
\end{equation}

The phenomenon of linear correlation or repeatability in the signal matrix can be seen as a correlation of coherent signals. This correlation can lead to information loss and redundancy in signal processing and analysis, limiting the accurate analysis and efficient utilization of signals. Such redundancy is considered a lack of rank in the matrix. To overcome this problem, considering that matrix summation can at least prevent the rank from decreasing, such methods often have the limitation of having strict fixed requirements for each matrix when dealing with high-dimensional data or complex signals. Therefore, the use of matrix summation aims to eliminate the correlation of coherent signals and improve the rank of the signal matrix.

\begin{equation}
	\sum_k\mathnormal{R_\mathnormal{X}^{(k)}} = A_1(\theta)(\sum_kB^{k-1}\mathnormal{R_\mathnormal{S}}(B^{k-1})^{H})A_1^H(\theta)+(\sum_k\sigma^2)\mathnormal{I}
\end{equation}

Assuming this addition is done $P$ times, where $P$ is referred to as the degree of spatial smoothing. Let's assume that $R_S$ is an extreme rank lacking matrix, that is, rank Matrix of ones. So it is necessary to continuously add to reduce the redundancy correlation and continuously unravel the coherence between signals. there is no harm in assuming $R_S=bb^H$, which can be referred to as the initial value in a certain sense, and the equation is written in the form of vector multiplication.

\begin{equation}
	\begin{split}
&\sum_{k=1}^PB^{k-1}\mathnormal{R_\mathnormal{S}}(B^H)^{k-1} \\
=&\sum_{k=1}^PB^{k-1}(bb^H)(B^H)^{k-1}\\
=&	\begin{pmatrix}
	b , Bb , B^2b , \cdots , B^{P-1}b\\
	\end{pmatrix}
	\begin{pmatrix}
	b^H \\ b^HB^H \\ b^H(B^H)^{2} \\ \vdots \\ b^H(B^H)^{P-1}
	\end{pmatrix}
  	\end{split}
\end{equation}
Let the initial value of the vector b and the dynamic feature matrix of every two adjacent sub-arrays be
\begin{equation}
	\begin{split}
		b =
		\begin{pmatrix}
			b_1 \\ b_2 \\ \vdots \\ b_m
		\end{pmatrix} , B =
	\begin{pmatrix}
		V_1 & & \\
		    & \ddots & \\
		    & & V_m
	\end{pmatrix}
	\end{split}
\end{equation}
Then the form of the correlation matrix after summing becomes
\begin{equation}
\begin{split}
	&\begin{pmatrix}
		b_1 & V_1b_1 & V_1^2b_1 & \cdots & V_1^{P-1}b_1\\
		\vdots & \vdots & \vdots & \ddots & \vdots \\
		b_m & V_mb_m & V_m^2b_m & \cdots & V_m^{P-1}b_m\\
	\end{pmatrix}\\
	=&
	\begin{pmatrix}
			b_1 & & \\
			& \ddots & \\
			& & b_m
	\end{pmatrix}
	\begin{pmatrix}
	1 & V_1 & V_1^2 & \cdots & V_1^{P-1}\\
	\vdots & \vdots & \vdots & \ddots & \vdots \\
	1 & V_m & V_m^2 & \cdots & V_m^{P-1}\\
\end{pmatrix}
\end{split}
\end{equation}

It can be found that such correlation matrix is the result of the multiplication of diagonal matrix and Vandermonde matrix. It is known that the diagonal matrix is full rank, and the rank of Vandermonde matrix is full rank when the elements in the second column are different. This is very obvious. So the result of smoothing is that the smoothing degree is equal to the rank of the correlation matrix, and as the smoothing degree increases, the rank will also reach the same maximum value as the number of incoming signals $m$.

\begin{equation}
	\tilde{R}_\mathnormal{X} = A_1(\theta)(\frac{1}{P}\sum_{k=1}^PB^{k-1}\mathnormal{R_\mathnormal{S}}(B^{H})^{k-1})A_1^H(\theta)+\sigma^2\mathnormal{I}
\end{equation}
If the dynamic features are set to the following exponential form:
\begin{equation}
	V_i = e^{a_i\Delta t}
\end{equation}
Where $a_i$ is a complex value, whose real part represents the trend of dynamic feature changes, and the imaginary part represents the frequency of this dynamic feature.

\begin{equation}
	\begin{split}
	a_i &= \frac{1}{\Delta t}ln(V_i)\\
	&= \frac{1}{\Delta t}(ln|V_i|+iarg(V_i))\\
	&= \frac{1}{\Delta t}ln|V_i|+\frac{i}{\Delta t}arg(V_i)
	\end{split}
\end{equation}

\begin{figure*}[htb]
	\centering
	\includegraphics[width=\linewidth]{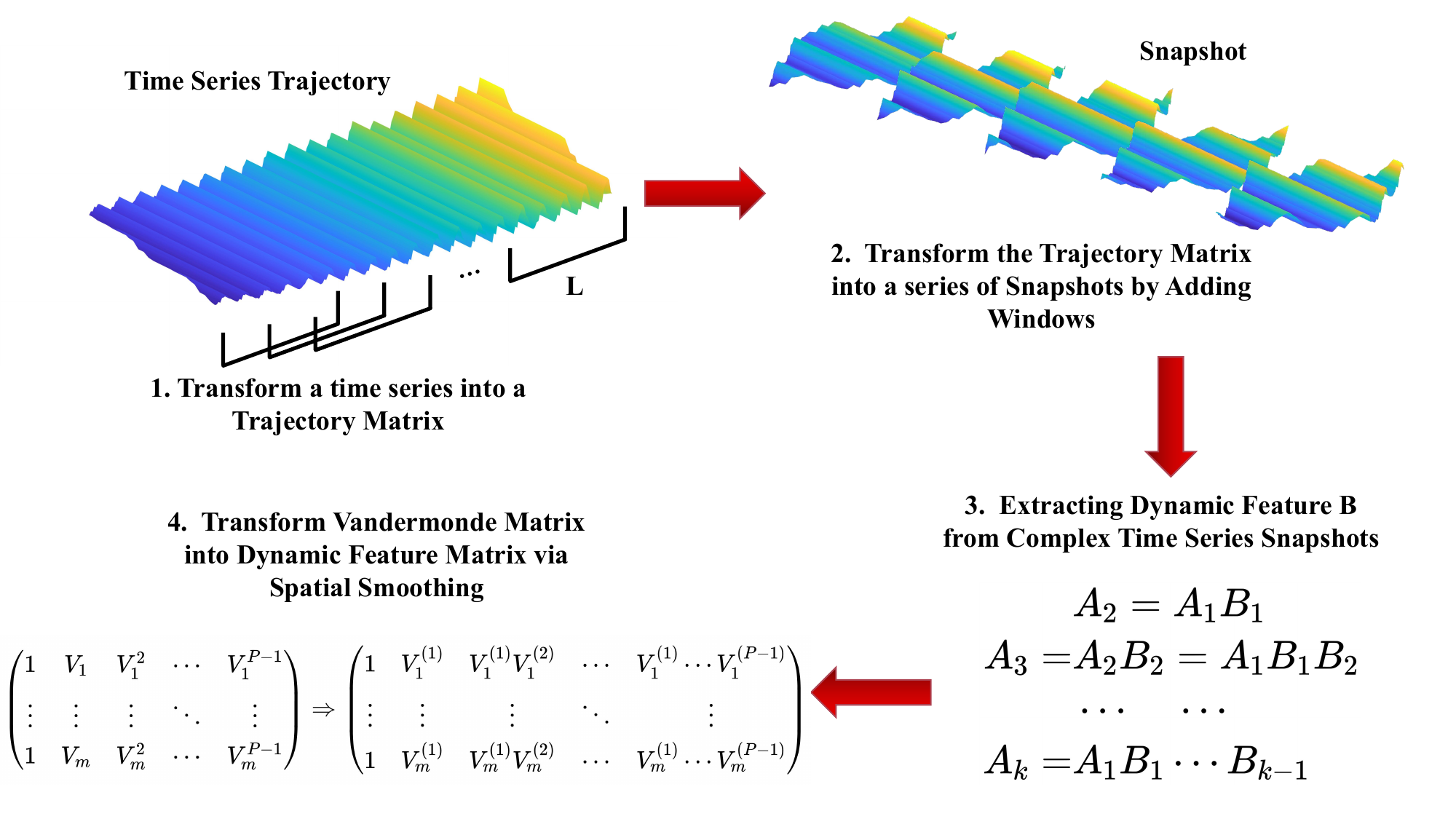}
	\caption{Schematic of data processing in Spatial Smoothing algorithm and Dynamic Features extraction.}
	\label{2}
\end{figure*}

It is worth mentioning that in array signal processing, the above order may be the case, but if it is a seasonal time series with a trend cycle, the order pattern in the Vandermonde matrix of the above situation may be destroyed. Instead, the row by row series representing the dynamic trend, and such series can be called dynamic features, which contain information about the change trend and frequency. This has an irreplaceable impact on our time series prediction. Fig. \ref{2} shows a schematic of the Spatial Smoothing algorithm and Dynamic Features extraction.

\subsection{Dynamic Feature Extraction}
Assuming a time series signal $x = x_1,x_2,\dots,x_n$, where n represents the signal data length. Through the Takens embedding theorem, it can be found that the topological equivalence method of time series for one-dimensional time series can construct a multidimensional time series matrix, namely the trajectory matrix. Therefore, the time series $x$ can construct the trajectory matrix $X$.

\begin{equation}
	X = 	
	\begin{pmatrix}
	x_1 & x_{1+\tau} & \cdots & x_{1+(d-1)\tau}\\
	x_2 & x_{2+\tau} & \cdots & x_{2+(d-1)\tau}\\
	\vdots & \vdots & \ddots & \vdots\\
	x_m & x_{m+\tau} & \cdots & x_{m+(d-1)\tau}\\	
	\end{pmatrix}
\end{equation}
Where $d$ represents the embedding dimension, tau represents the delay time, and $m=n - (d-1)\tau$. In the trajectory matrix $X$, there are two main parameters $d$ and $\tau$, and different d and tau can construct different matrices $X$, which have a significant impact on the analysis results. 
\begin{equation}
	\mathnormal{X} = A(\theta)\mathnormal{S} + \mathnormal{N}
\end{equation}
Where $\theta = (\theta_1,\cdots,\theta_m)$ can represent the changes in data in the time series trajectory matrix. In array processing, subspace theory will be widely applied. Here we briefly elaborate on the important conclusions of the subspace. Assuming we can represent the entire complex field using $r$ subspaces.
\begin{equation}
	\mathnormal{C}^n = U_1 \oplus U_2 \oplus \cdots \oplus U_r
\end{equation}

If we want to prove that the subspace $A_1$ and $A_2$ are one subspace, we have the following steps:
\begin{equation}
	\begin{split}
	\mathnormal{C}^n = Span(A) \oplus ker(A^H)\\
	ker(A^H) = \{x:A^Hx = 0\}\\
	ker(A_1^H) = ker(A_2^H)\\
	\Longrightarrow Span(A_1) = Span(A_2)
	\end{split}
\end{equation}

It is necessary to make a correlation matrix for the trajectory matrix $X$ of the time series and use PCA to reduce its dimensionality.
\begin{equation}
	\begin{split}
	\mathnormal{R_\mathnormal{X}} =& \mathnormal{E}(\mathnormal{X}\mathnormal{X}^H)\\
	=& \sum^n_{k=1} \lambda_k U_k U_k^H\\
	=&
	\begin{pmatrix}
		U_S & U_N
	\end{pmatrix}
	\begin{pmatrix}
		\Lambda_S & \\
		& \Lambda_N
	\end{pmatrix}
\begin{pmatrix}
	U_S^H \\ \quad\\ U_N^H
\end{pmatrix}
	\end{split}
\end{equation}

Obviously, for the universal matrix $X$, it can be divided into signal subspace and noise subspace. The diagonal matrix lambda values in the signal subspace are generally greater than zero, while the diagonal matrix values in the noise subspace are very insignificant. I.e. 
$\lambda_1 \geq \lambda_2 \geq \cdots \geq \lambda_n \geq 0$. Where $U_S = (U_1,\cdots,U_m)$ is the Signal Subspace and $U_N=(U_{m+1},\cdots,U_n)$ is the Noise Subspace.
So naturally, under the assumption of white noise, the correlation matrix can be written in the following form:

\begin{equation}
	\begin{split}
		\mathnormal{R_\mathnormal{X}}
		&=A(\theta)\mathnormal{R_\mathnormal{S}}A^H(\theta)+\mathnormal{E}(\mathnormal{N}\mathnormal{N}^H)\\
		&=A(\theta)\mathnormal{R_\mathnormal{S}}A^H(\theta)+\sigma^2\mathnormal{I}
	\end{split}
\end{equation}

After organizing and comparing the results of PCA with the calculation results of the correlation matrix, it can be concluded that:
\begin{equation}
	\begin{split}
			\begin{pmatrix}
			U_S & U_N
		\end{pmatrix}
		&\begin{pmatrix}
			\Lambda_S & \\
			& \Lambda_N
		\end{pmatrix}
		\begin{pmatrix}
			U_S^H \\ \quad\\ U_N^H
		\end{pmatrix}
		=A(\theta)\mathnormal{R_\mathnormal{S}}A^H(\theta) + \sigma^2\mathnormal{I}\\
		\Longrightarrow  
		\begin{pmatrix}
			U_S & U_N
		\end{pmatrix}
		&\begin{pmatrix}
			\Lambda_S - \sigma^2\mathnormal{I} & \\
			& \Lambda_N - \sigma^2\mathnormal{I}
		\end{pmatrix}
		\begin{pmatrix}
			U_S^H \\ \quad\\ U_N^H
		\end{pmatrix}
		=A(\theta)\mathnormal{R_\mathnormal{S}}A^H(\theta)\\
		\Longrightarrow 
		&U_S(\Lambda_S - \sigma^2 \mathnormal{I})U_S^H = A(\theta)R_SA^H(\theta)
	\end{split}
\end{equation}

Through observation, it can be found that the relationship between $U_S$ and $A(\theta)$ is highly correlated. We may be able to prove through the steps of equation $(18)$ that the subspace formed by $U_S$ and $A(\theta)$ is the same subspace.

\begin{equation}
	Span(U_S) = Span(A(\theta))
\end{equation}
Proof:
\begin{equation}
	\begin{split}
		&\forall x \in ker(U_S^H) , \quad U_S^Hx=0\\
		\Longrightarrow &U_S(\Lambda_S - \sigma^2I)U_S^Hx = A(\theta)R_SA^H(\theta)x = 0\\
		\Longrightarrow & x^HA(\theta)R_SA^H(\theta)x = 0\\
		\Longrightarrow &
		(A^H(\theta)x)^HR_S(A^H(\theta)x) = 0\\
		\Longrightarrow &
		A^H(\theta)x = 0\\
		\Longrightarrow &	
		x \in ker(A^H(\theta))
	\end{split}
\end{equation}
So for $ X \in \mathnormal{C}^N $, $S \in \mathnormal{C}^M $, $M < N$, use the same method to make the correlation matrix:
\begin{equation}
	\begin{split}
		\mathnormal{R_\mathnormal{X}} =& \mathnormal{E}(\mathnormal{X}\mathnormal{X}^H)\\
		=& \sum^n_{k=1} \lambda_k U_k U_k^H\\
		=&A_k(\theta)\mathnormal{R_\mathnormal{S}}A_k^H(\theta)+\sigma^2\mathnormal{I}\\
		=&U_S\Lambda_SU_S^H + U_N\Lambda_NU_N^H
	\end{split}
\end{equation}

Since it has been proven that $U_S$ and $A(\theta)$ are the same subspace, it can be considered that there is a rotation matrix $T$ without scale transformation between them:
\begin{equation}
Span(U_S) = Span(A(\theta))\Longrightarrow A(\theta)T = U_S
\end{equation}
Where $T$ can be called rotation invariant transform matrix. Next, we will extract the dynamic features we need using the ESPRIT algorithm created by Roy in the array signal processing algorithm. This algorithm requires dividing both subspace into upper and lower blocks in the same two ways:
\begin{equation}
	\begin{split}
	A(\theta) =& 
	\begin{pmatrix}
		A_1(\theta)\\
		*
	\end{pmatrix}
	=
	\begin{pmatrix}
		*\\
		A_2(\theta)
	\end{pmatrix}\\
		U_S(\theta) = &
	\begin{pmatrix}
		U_S^{(1)}(\theta)\\
		*
	\end{pmatrix}
	=
	\begin{pmatrix}
		*\\
		U_S^{(2)}(\theta)
	\end{pmatrix}
	\end{split}
\end{equation}

After analyzing $A(\theta)$ in the previous text, it can be found that there is a certain phase difference relationship between $A_1(\theta)$ and $A_2(\theta)$, which is caused by the assumed delay when creating the trajectory array. Their relationship is as follows, and the relationship between each of them can be further inferred.
\begin{equation}
	\begin{split}
		\left\{ 
		\begin{aligned}
			&A_2(\theta) = A_1(\theta)\Psi\\
			& A(\theta)T = U_S \\
			&\Psi = diag(e^{j\phi_1},\cdots,e^{j\phi_m})
		\end{aligned}
		\right.
		\Longrightarrow&\left\{ 
		\begin{aligned}
			&A_1(\theta)T = U_S^{(1)}\\
			&A_2(\theta)T = U_S^{(2)} \\
			&A_2(\theta) = A_1(\theta)\Psi\\	
			&A_1(\theta) = U_S^{(1)}T^{-1}\\
			&A_2(\theta) = U_S^{(2)}T^{-1}
		\end{aligned}
	\right.
	\end{split}
\end{equation}

Due to the fact that the matrix $T$ is an orthogonal matrix with rotational invariant transform, the specific $T$ can be obtained through pseudo inverse avoidance to obtain the information carried by $T$, as $\Psi$ and its congruent transformation are similar matrices.
\begin{equation}
	\begin{split}
	&U_S^{(2)}T{-1} = U_S^{(1)}T^{-1}\Psi\\
	\Longrightarrow&
	U_S^{(2)} = U_S^{(1)}T^{-1}\Psi T\\
	\Longrightarrow&
	(U_S^{(1)})^{\dag}U_S^{(2)} = T^{-1}\Psi T\\
	\Longrightarrow&
	((U_S^{(1)})^{H}U_S^{(1)})^{(-1)}(U_S^{(1)})^{H}U_S^{(2)} = T^{-1}\Psi T\sim\Psi\\
	\end{split}	
\end{equation}

Due to the addition of windows in the spatial smoothing algorithm before, there are many such $\Psi$. We arrange the diagonal values of each $\phi$ in a sequence, which we call dynamic features. Afterwards, it can be predicted using neural networks.

\subsection{General Variational Mode Decomposition (GVMD) and Sequential General Variational Mode Decomposition (SGVMD)}
In order to extend the applicability of the VMD algorithm, the GVMD algorithm is considered. Essentially, it is a generalization and eigenvalueization of the VMD algorithm, which extends the variational mode decomposition method to the general field. The general form of the loss function of the algorithm is shown in equation \ref{29}:
\begin{equation} 
	\begin{split}
		\mathcal{L}\left(\left\{\hat{u}_{i}\right\}\right)=\left\|\hat{f}-\sum_{i=1}^{K} \hat{u}_{i}\right\|^{2}+\alpha \sum_{i=1}^{K}\left\|g\left(\hat{u}_{i}\right)\right\|^{2} 
	\end{split}
	\label{29}
\end{equation}
Where the function $g\left(\hat{u}_{i}\right)$ represents the characteristic feature of ${\hat{u}}_{i}$.

VMD is just a special case of GVMD when the components have narrowband characteristics in the frequency domain. GVMD is more general and not limited to the time and frequency domains, nor does it have fixed requirements for component characteristics.

However, both VMD and GVMD algorithms have a common drawback, which is the need to know the modal number in advance. In practical situations, it is often difficult to obtain this information. Therefore, it is necessary to introduce a regularization approach to GVMD, namely the SGVMD algorithm, to achieve the sequential extraction of each modal component of a time series in the case of unknown modal numbers.

The SGVMD algorithm is not sensitive to the initial values of the modes. The output order of the components is also flexible, and adjusting the initial values can change the extraction order. The convergence speed of the algorithm is also fast. In order to reduce the number of extraction iterations and further accelerate the convergence process, the initial peak of the mode can be set to the spectral peak of the current residual mixed component. Existing decomposition algorithms are highly dependent on the modal number, but the SGVMD algorithm overcomes this difficulty, which greatly extends its application scope. The SGVMD algorithm splits the mixed mode into two parts each time. In order to achieve the extraction of the current mode, the loss function established imposes stricter constraints on the current mode.

Taking non-stationary time series decomposition as an example. Firstly, we need to consider maximizing the fidelity between the extracted components and the current remaining components. Because maximizing fidelity can be achieved by minimizing residual terms, component fidelity terms can be designed as:
\begin{equation} 
		\| \hat{f}_{i-1}^r(\omega) - \hat{u}_i(\omega) \|_2 
\end{equation}
Where $\hat{f}_{i-1}^r$ is the remaining part of the mixed sequence after the $(i-1)$th extraction

In order to describe narrowband characteristics in the frequency domain, the introduction of current component feature terms is also necessary. This item can be represented as:
\begin{equation} 
		\| \hat{u}_{i}(\omega) (\omega - \omega_{c,i}) \|_2 
	\label{31}
\end{equation}

The above two constraints can only constrain the current mode. If we do not impose any restrictions on the remaining modes, we still cannot achieve the purpose of sequential extraction. Therefore, in order to limit the remaining modes to narrowband components, we should introduce a third term, which is expressed as follows:
\begin{equation} 
		\| \hat{f}_{i}^r(\omega) (\omega - \omega_{c,i}^r) \|_2 
	\label{32}
\end{equation}

In conclusion, when the component features of a time series can be defined by explicit feature terms, SGVMD can effectively solve its sequential decomposition problem. The loss function \ref{31} and \ref{32} introduces two penalty factors, $\alpha$ and $\beta$, but experiments have shown that the ratio between the two factors is what affects the separation results. Therefore, simplifying the equation by setting one of the factors to 1 reduces the number of parameters needed.

the loss function of the SGVMD algorithm can be rewritten as Equation \ref{33}:
\begin{equation} 
	\begin{aligned}
		&\mathcal{L}\left( {\hat{u}}_{i} \right) = \left\| {{\hat{f}}_{i - 1}^{r}(\omega) - {\hat{u}}_{i}(\omega)} \right\|_{2}^{2} + \alpha\left\| {{\hat{u}}_{i}(\omega)\left( {\omega - \omega_{c,i}} \right)} \right\|_{2}^{2}\\
		&+ \beta\left\| {{\hat{f}}_{i}^{r}(\omega)\left( {\omega - \omega_{c,i}^{r}} \right)} \right\|_{2}^{2}
	\end{aligned}
	\label{33}
\end{equation}

The process of SGVMD algorithm applied to non-stationary time series decomposition is shown in Algorithm \ref{alg1}:
\begin{algorithm}[H]
	\caption{SGVMD algorithm for non-stationary time series.}\label{alg:alg1}
	\begin{algorithmic}[1]
		\STATE {\textsc{INPUT:time series dataset }$Y = \{ y_1,y_2,\cdots,y_n\};$}
		\STATE {\textsc{OUTPUT:Mode components }$\hat{u_1},\hat{u_2},\cdots,\hat{u_i};$}
		\STATE \hspace{0.5cm}$ \textbf{Initialize } \epsilon , \alpha  \textbf{ and }\beta  ,\textbf{ let } \hat{f}_0^r(\omega) = \hat{f}(\omega);$
		\STATE \hspace{0.5cm}$ \textbf{Let } i = 1;$
		\STATE \hspace{0.5cm}$ \textbf{WHILE } \left\| {{\hat{f}}_{i}^{r}(\omega)} \right\|_{2}^{2} > \epsilon \textbf{ DO }$
		\STATE \hspace{0.5cm}\hspace{0.5cm}$ \textbf{Initialize } \hat{u}_i^0(\omega) \textbf{ and } \eta;$
		\STATE \hspace{0.5cm}\hspace{0.5cm}$ \textbf{Let } s = 1;$
		\STATE \hspace{0.5cm}\hspace{0.5cm}$ \textbf{DO}$
		\STATE \hspace{0.5cm}\hspace{0.5cm}\hspace{0.5cm}$ \omega_{c,i}^{s-1}=\frac{\int^{+\infty}_0 \omega | \hat{u}_i^{s-1}(\omega) |^2 d\omega}{\int^{+\infty}_0  | \hat{u}_i^{s-1}(\omega) |^2 d\omega} ;$
		\STATE \hspace{0.5cm}\hspace{0.5cm}\hspace{0.5cm}$ \hat{f}_i^{r,s-1}(\omega) = \hat{f}_{i-1}^r(\omega) - \hat{u}_i^{s-1}(\omega);$
		\STATE \hspace{0.5cm}\hspace{0.5cm}\hspace{0.5cm}$ \omega_{c,i}^{r,s-1}=\frac{\int^{+\infty}_0 \omega | \hat{f}_i^{r,s-1}(\omega) |^2 d\omega}{\int^{+\infty}_0  | \hat{f}_i^{r,s-1}(\omega) |^2 d\omega} ;$
		\STATE \hspace{0.5cm}\hspace{0.5cm}\hspace{0.5cm}$ \hat{u}_i^s(\omega) = \frac{ \hat{f}_{i-1}^r(\omega)(1+\beta (\omega - \omega_{c,i}^{r,s-1})^2) }{1 + \alpha (\omega - \omega_{c,i}^{s-1})^2 + \beta(\omega - \omega_{c,i}^{r,s-1})^2}; $
		\STATE \hspace{0.5cm}\hspace{0.5cm}\hspace{0.5cm}$ s = s + 1; $
		\STATE \hspace{0.5cm}\hspace{0.5cm}$\textbf{WHILE } \| \hat{u}_i^{s-1}(\omega) - \hat{u}_i^{s-2}(\omega) \|_2^2 > \eta $
		\STATE \hspace{0.5cm}\hspace{0.5cm}$ \hat{u}_i(\omega) = \hat{u}_i^s(\omega); $
		\STATE \hspace{0.5cm}\hspace{0.5cm}$\hat{f}_i^r(\omega) = \hat{f}_{i-1}^r(\omega) - \hat{u}_i(\omega); $
		\STATE \hspace{0.5cm}\hspace{0.5cm}$i = i + 1;$
		\STATE \hspace{0.5cm}$\textbf{END WHILE }$
	\end{algorithmic}
	\label{alg1}
\end{algorithm}

When the component features of a time series can be defined by explicit feature terms, SGVMD can effectively solve its sequential decomposition problem. The loss function equation \ref{33} introduces two penalty factors $\alpha$ and $\beta$, but experiments have found that the ratio of the two factors affects the separation results. Therefore, simplifying it by setting one of them to 1 reduces the number of parameters.

\subsection{Long Short-Term Memory Network}

Recurrent Neural Networks (RNNs) are a class of neural networks that excel in processing sequential data. However, traditional RNNs suffer from vanishing or exploding gradients when dealing with long sequences, making it difficult to capture long-term dependencies. LSTM, as an improved RNN architecture, overcomes these issues through sophisticated memory cells and gating mechanisms, making it an essential choice for sequence data processing.

The core component of an LSTM network is the LSTM cell. A typical LSTM cell consists of three essential parts: the forget gate, the input gate, and the output gate. These gates control the flow of information, enabling LSTM to selectively forget, read, and output information.

\begin{figure}[H]
	\centering
	\includegraphics[width=\linewidth]{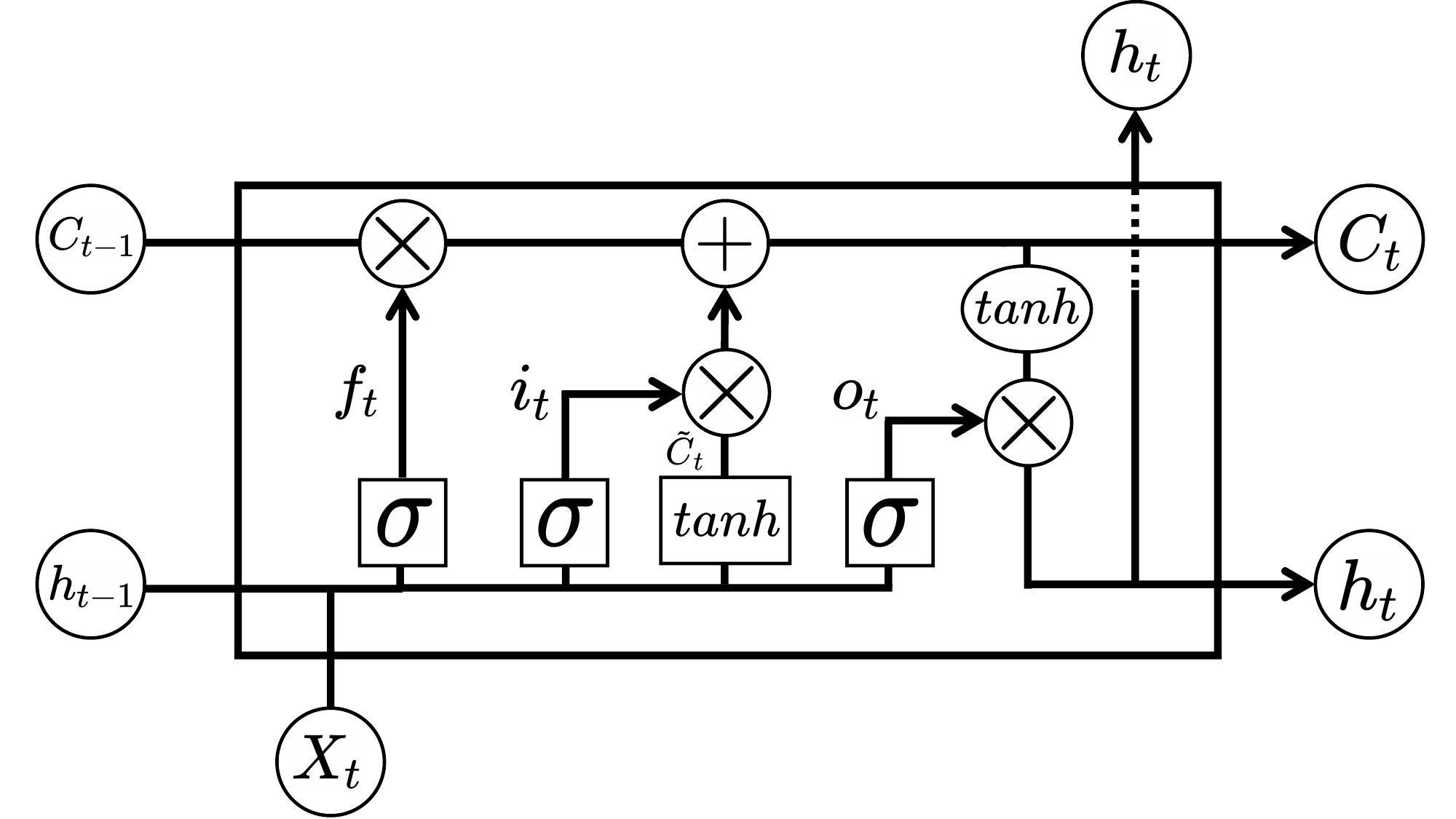}
	\caption{The LSTM cell}
	\label{3}
\end{figure}

"The LSTM Cell" below presents a typical LSTM unit structure. Fig. \ref{3} illustrates the key components inside the LSTM cell and how information is propagated. Please note that the specific parameters and weight values in the figure may vary depending on the task and model training in practical applications.

LSTM processes input sequences through a series of time steps. At each time step $t$, LSTM receives the current input $x(t)$ and the previous hidden state $h(t-1)$ as inputs and performs the following steps for information processing:

Forget Gate: The forget gate determines which parts of the previous hidden state $h(t-1)$ should be forgotten, with an output value $f(t)\in[0, 1]$.
\begin{equation}
	f_t = \sigma (W_f \cdot [h_{t-1},x_t] + b_f)
\end{equation}

Input Gate: The input gate decides which information to include in the memory based on the current input $x(t)$ and the previous hidden state $h(t-1)$, producing an output value $i(t)\in[0, 1]$.
\begin{equation}
	C_t = f_t \times C_{t-1} + i_t \times \tilde{C}_t
\end{equation}

Update Memory: Using the forget gate $f(t)$ and the input gate $i(t)$, LSTM updates the previous hidden state $h(t-1)$ to obtain the current candidate memory $\tilde{C}(t)$.
\begin{equation}
	\begin{split}
	i_t &= \sigma (W_i \cdot [h_{t-1},x_t] + b_i)\\
	\tilde{C}_t =& tanh(W_C \cdot [h_{t-1},x_t] + b_C)
	\end{split}
\end{equation}

Output Gate: The output gate, considering the current input $x(t)$, the previous hidden state $h(t-1)$, and the updated memory $\tilde{C}(t)$, determines the current hidden state $h(t)$ and output $y(t)$.
\begin{equation}
	\begin{split}
		o_t = \sigma &(W_o \cdot [h_{t-1},x_t] + b_o)\\
		h_t &= o_t \cdot tanh(C_t)
	\end{split}
\end{equation}

Training strategy refers to combining multiple models, and each extracted dynamic requires an LSTM neural network model. This method indicates that neural networks learn general features of similar patterns between different training datasets. The implicit assumption of this method is that the changes in each dynamic feature sequence are similar to each other and can perform well in LSTM. However, the dynamic features obtained during the spatial smoothing process mentioned earlier are a linear combination of the eigenvalues in the initial diagonal matrix and the dynamic feature sequence in the dynamic feature matrix. The interaction between the observed patterns and nonlinearity in the experiment cannot be clearly explained. Therefore, in reality, different dynamic feature sequences cannot be artificially limited to the same LSTM prediction model parameters, and changes in dynamic features may be distributed in a wide Gaussian distribution or a certain oscillation frequency. Therefore, the single parameter model method cannot solve all the extracted dynamic complete features, as it only learns from similar training patterns. Another challenge of the multi parameter method is to increase the memory required to embed LSTM models in electronic hardware. Utilizing more modalities to improve the accuracy of reduced order models can lead to memory bottlenecks in modalities.
\begin{figure*}[htb]
	\centering
	\includegraphics[width=\linewidth]{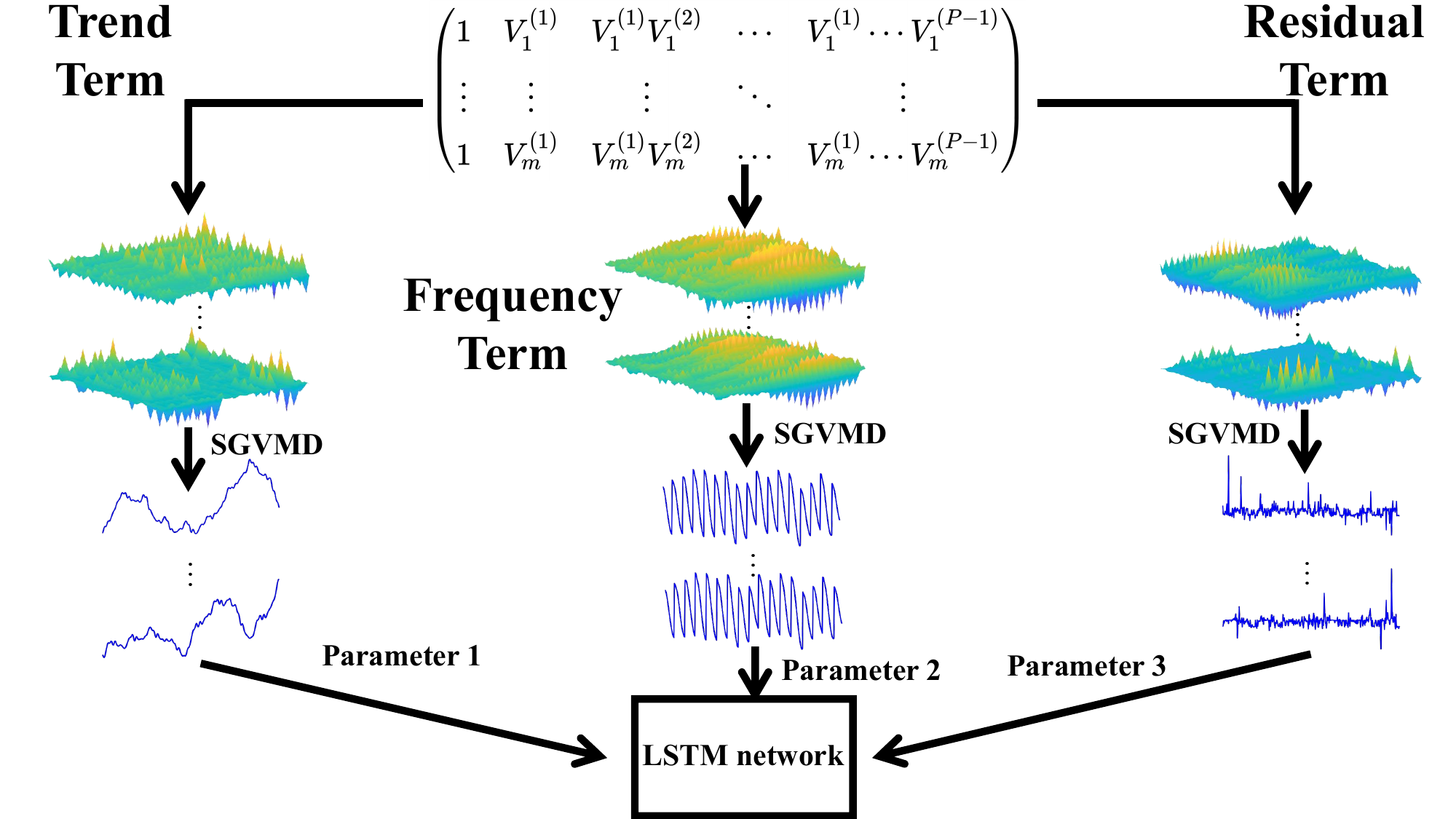}
	\caption{After the dynamic feature extraction is completed, it is divided into three different types according to the variation form of the dynamic feature series via SGVMD and the workflow diagram of the multi-parameter LSTM method.}
	\label{4}
\end{figure*}

\section{Proposed Combination Method}
\subsection{Datasets}

\begin{figure}[H]
	\centering
	\includegraphics[width=\linewidth]{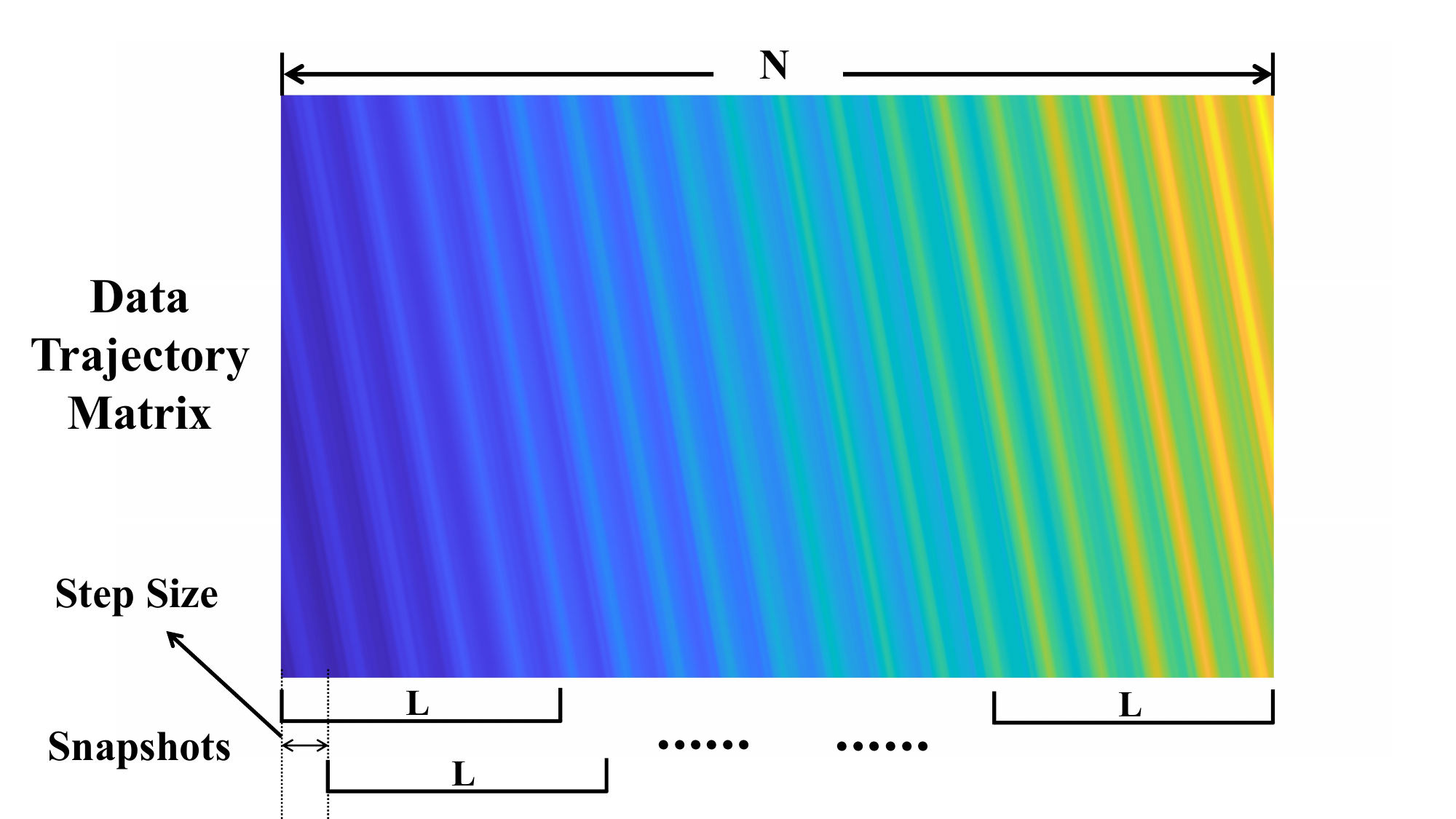}
	\caption{Construction of snapshots from the data trajectory matrix.}
	\label{5}
\end{figure}

In time series analysis, the data sample trajectory matrix constructed through the Takens embedding theorem is usually segmented and constructed using sliding time window technology. This article constructs data samples separately for different parameters of spatial smoothing algorithms and LSTM models. Fig. \ref{5} shows the construction process of the data snapshot matrix. The spatial smoothing algorithm extracts dynamic features from a non-stationary, seasonal, and periodic trend set of market product sales data. Assume the window length $L$ and sliding step is set to 1, and the source data is cut through a sliding time window to obtain the data snapshot matrix $X$ required by the spatial smoothing algorithm.

\begin{figure}[H]
	\centering
	\includegraphics[width=\linewidth]{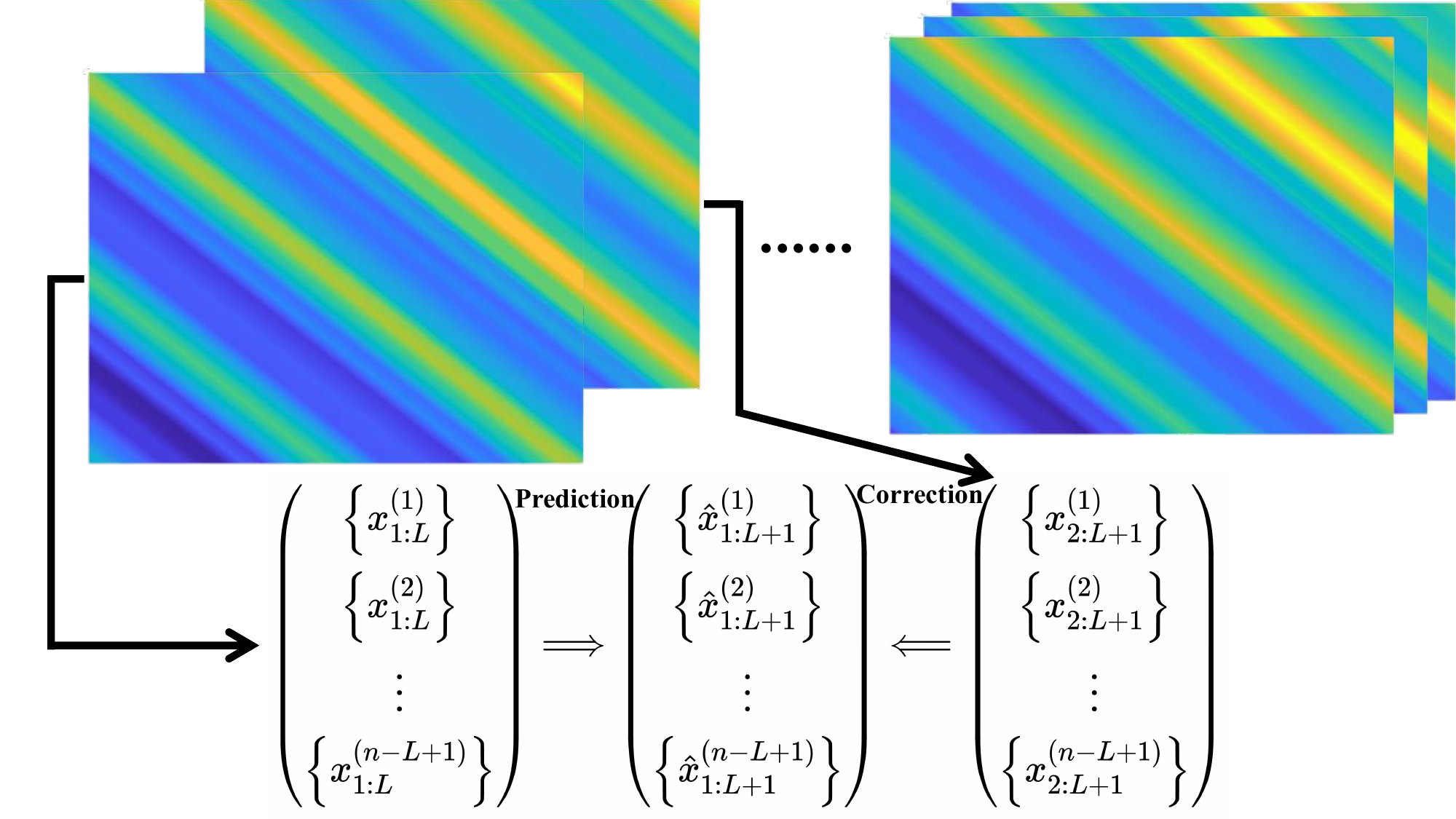}
	\caption{Construction of dynamic feature matrix and its prediction.}
	\label{6}
\end{figure}

The data modeled by the LSTM model is the dynamic feature sequence of each snapshot data. As shown in Fig. \ref{6}, set the width of each snapshot to L, intercept each data unit with a step size of 1 from the snapshot set to construct a sample, and take the sales volume on the day after the window as the prediction target. By continuously moving backwards, a series of overlapping sample data is formed, which reduces the time-varying characteristics of the data while utilizing the temporal information of the data. The resulting dataset has significant differences in the range of different eigenvalues, which is not conducive to the convergence of the LSTM model. It is also necessary to use different parameters to address the characteristics of different LSTM datasets and perform normalization operations to adjust the data distribution.

\subsection{Dynamic Feature Extraction and Prediction}
\subsubsection{Small amplitude oscillation trend seasonal periodic time series}
The data processing method of spatial smoothing algorithm is suitable for complex system analysis, but its application in market sales prediction assumes that the market is a complex system with overall correlation, that is, sales fluctuations during a certain period may be influenced by itself and relevant industry sectors. On this basis, spatial smoothing is used to extract dynamic features representing the overall market trend changes and price changes over the past n days, capturing the essential characteristics and potential behavioral patterns of sales changes. The application object of spatial smoothing algorithm in empirical research comes from the sales volume of a certain online store from May 5th, 2000 to April 6th, 2003.

The dynamic characteristics of the whole system can be obtained by extracting the dynamic characteristics of the data snapshot matrix $X$ using spatial smoothing, which is expressed in the form similar to the Vandermonde matrix. The decomposition order of the initial value $b$ corresponds to the initial value size of the market dynamic feature information. The first row of dynamic feature value sequence $X_{1:L}^{(1)}$ obtained from the system decomposition is called the dominant feature, which can also be called the trend component, representing the main trend information of the entire market. Due to the fact that the values of dynamic features are generally complex, the real part represents the trend of dynamic feature changes, and the imaginary part represents its frequency. And if its modulus is greater than 1, it means that the market has an expansion trend; Its modulus is less than 1, indicating a tightening trend in the market; When the modulus is equal to 1, it indicates that the overall market trend is stable. Especially for eigenvalues without imaginary parts, the corresponding change rate of Market trend is exponential, so the influence of this mode on the system is more significant than the dynamic characteristics of other components, and it is one of the important factors leading to market changes.

\begin{figure*}[htb]
	\centering
	\subfigure[]{
		\includegraphics[scale=0.25]{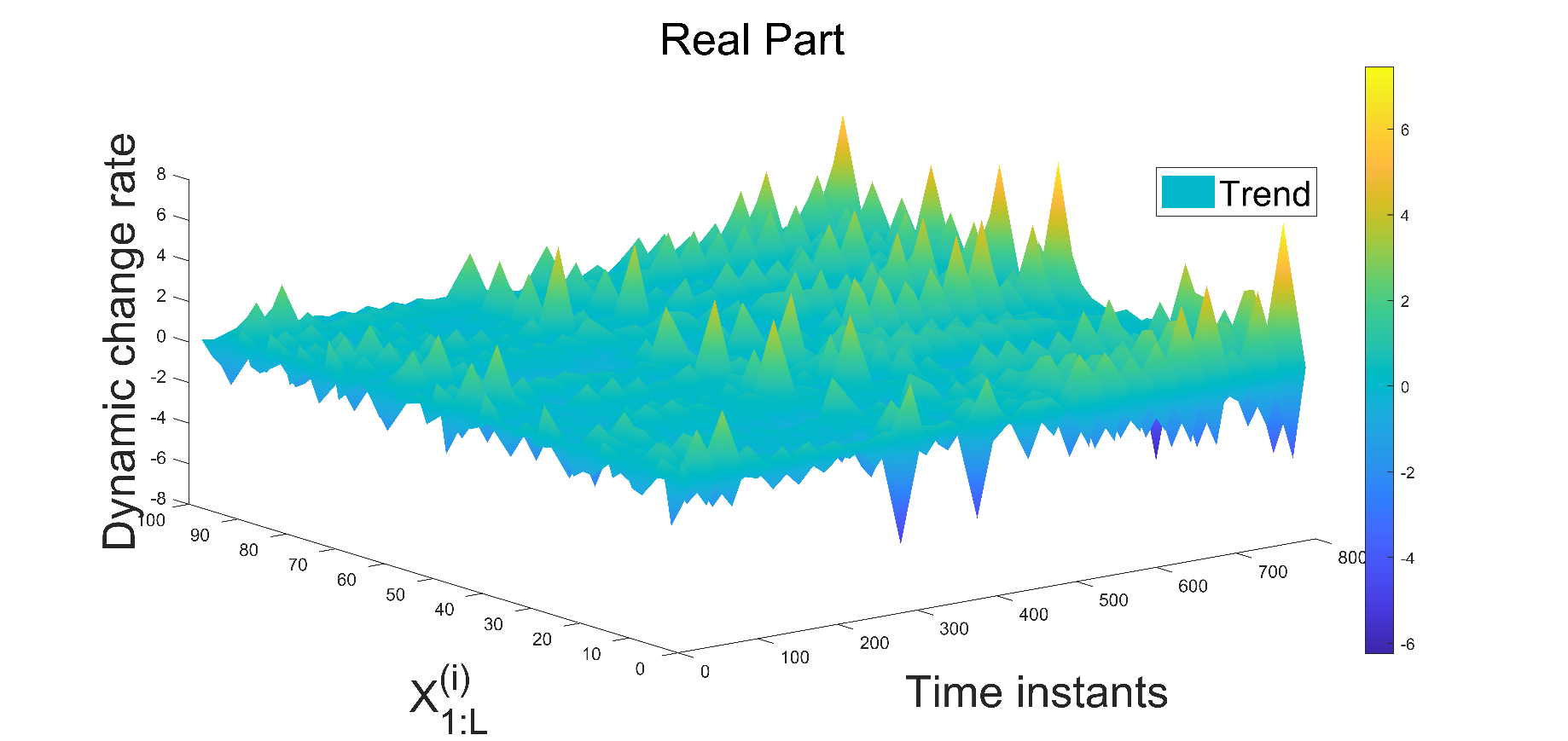}}
		\label{a1}
	\subfigure[]{
		\includegraphics[scale=0.25]{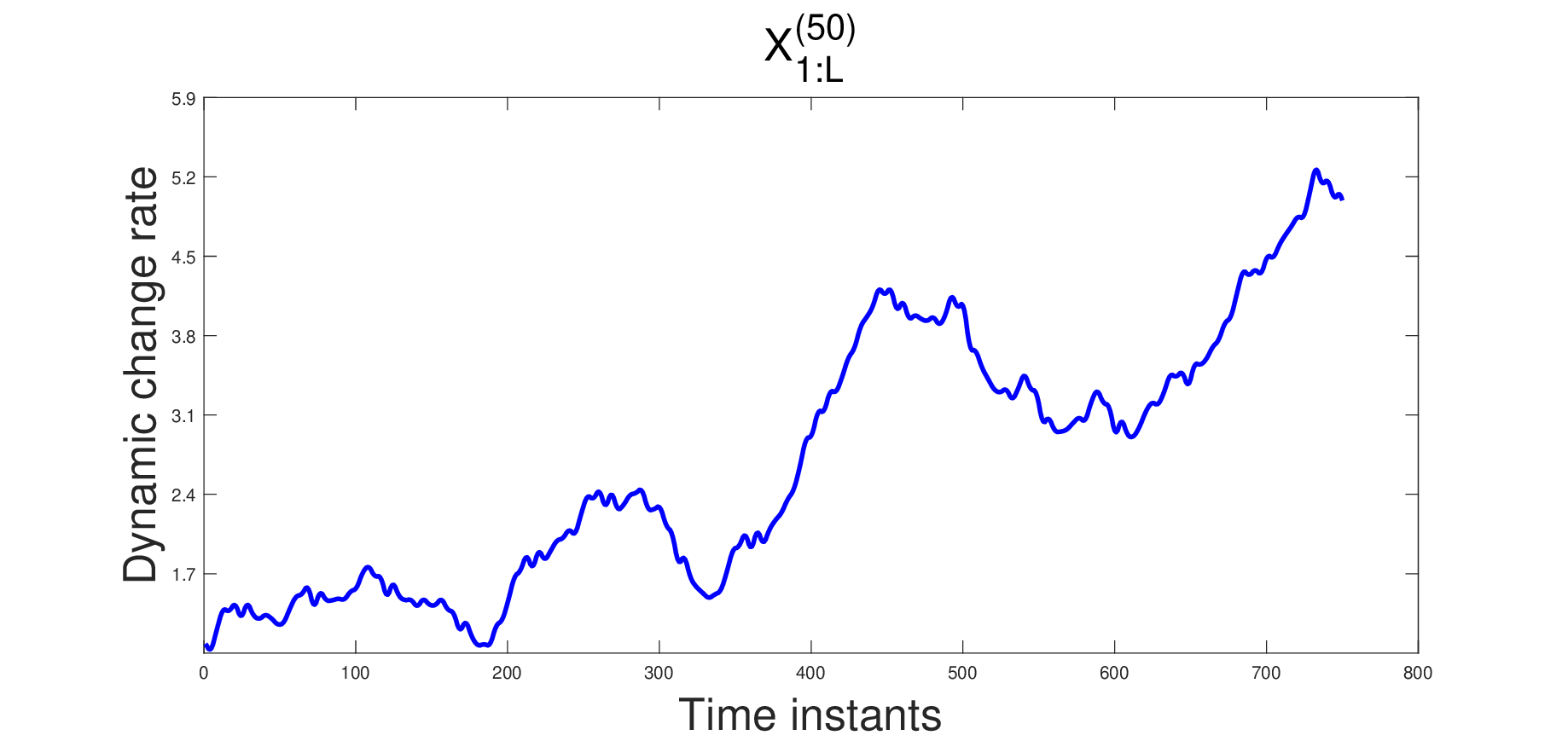}}
		\label{b1}
	\\
	\subfigure[]{
		\includegraphics[scale=0.25]{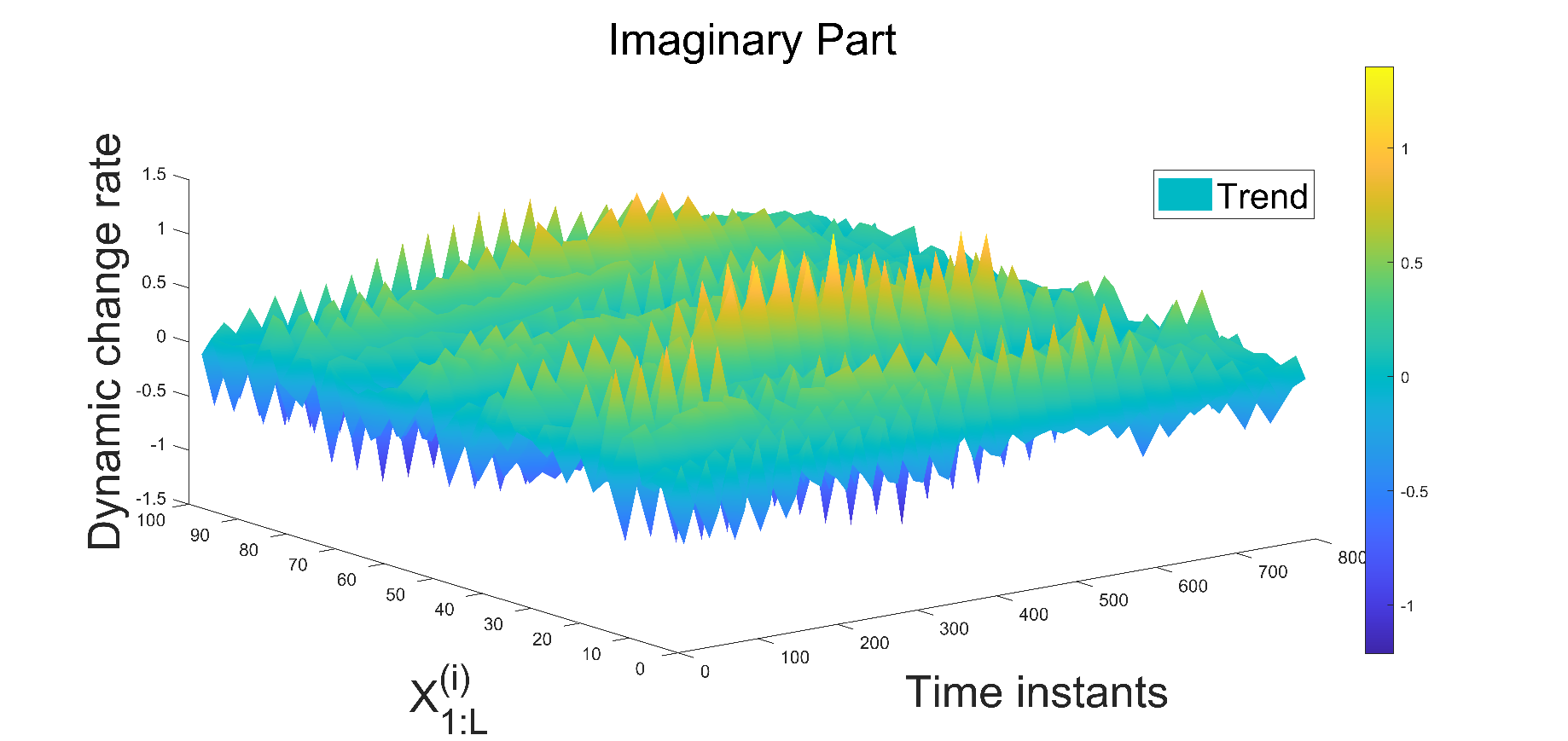}}
		\label{c1}
	\subfigure[]{
		\includegraphics[scale=0.25]{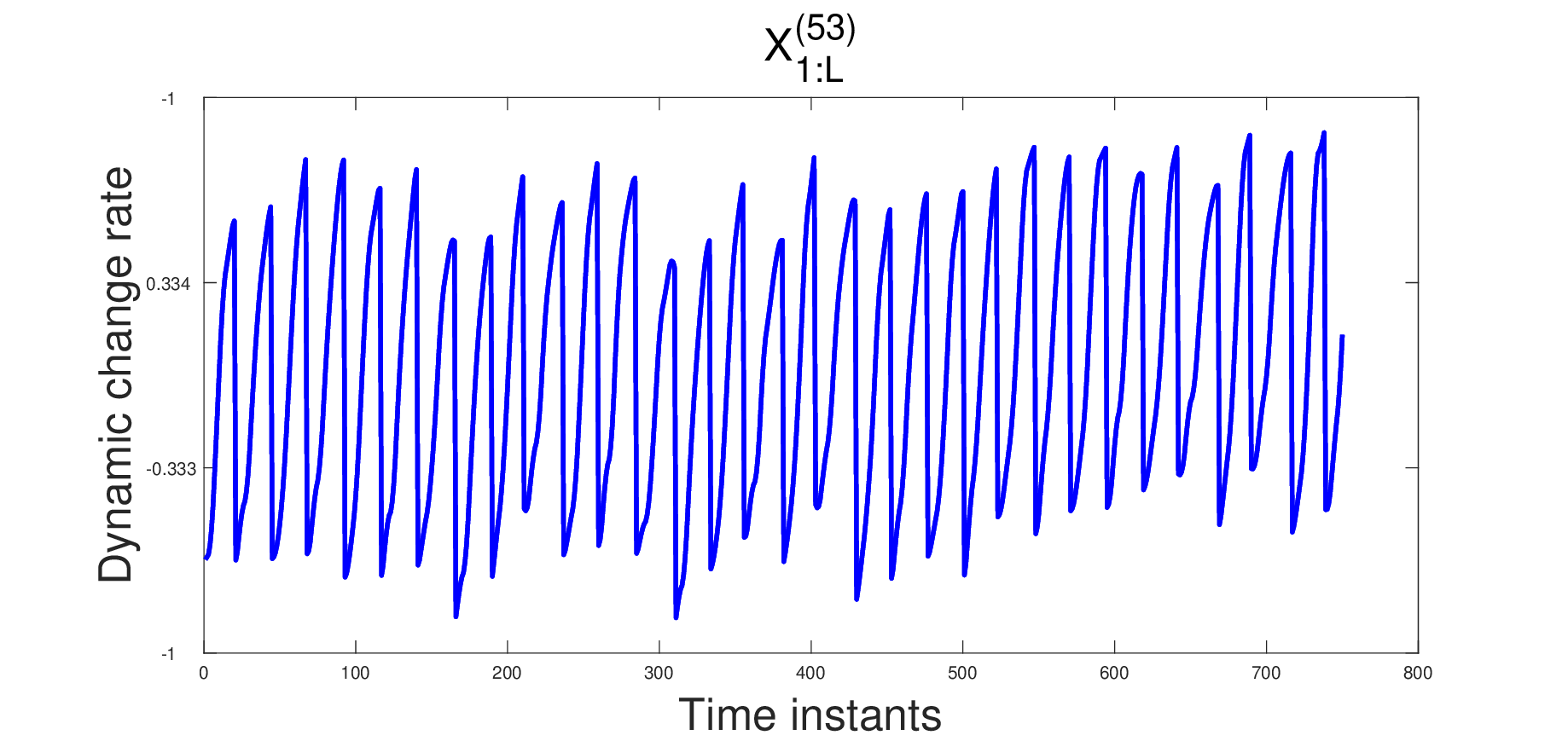}}
		\label{d1}
	\\
		\subfigure[]{
		\includegraphics[scale=0.25]{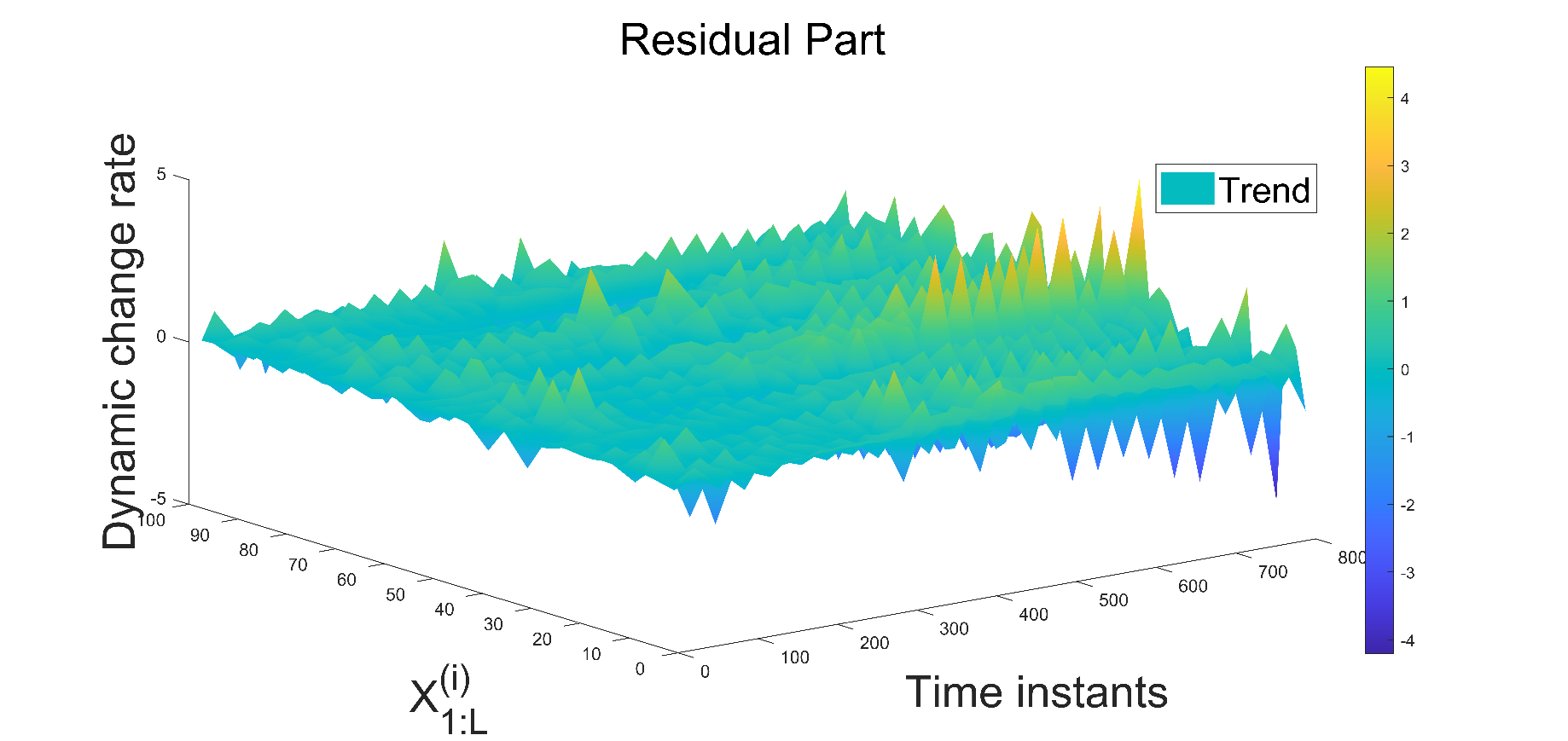}}
		\label{e1}
	\subfigure[]{
		\includegraphics[scale=0.25]{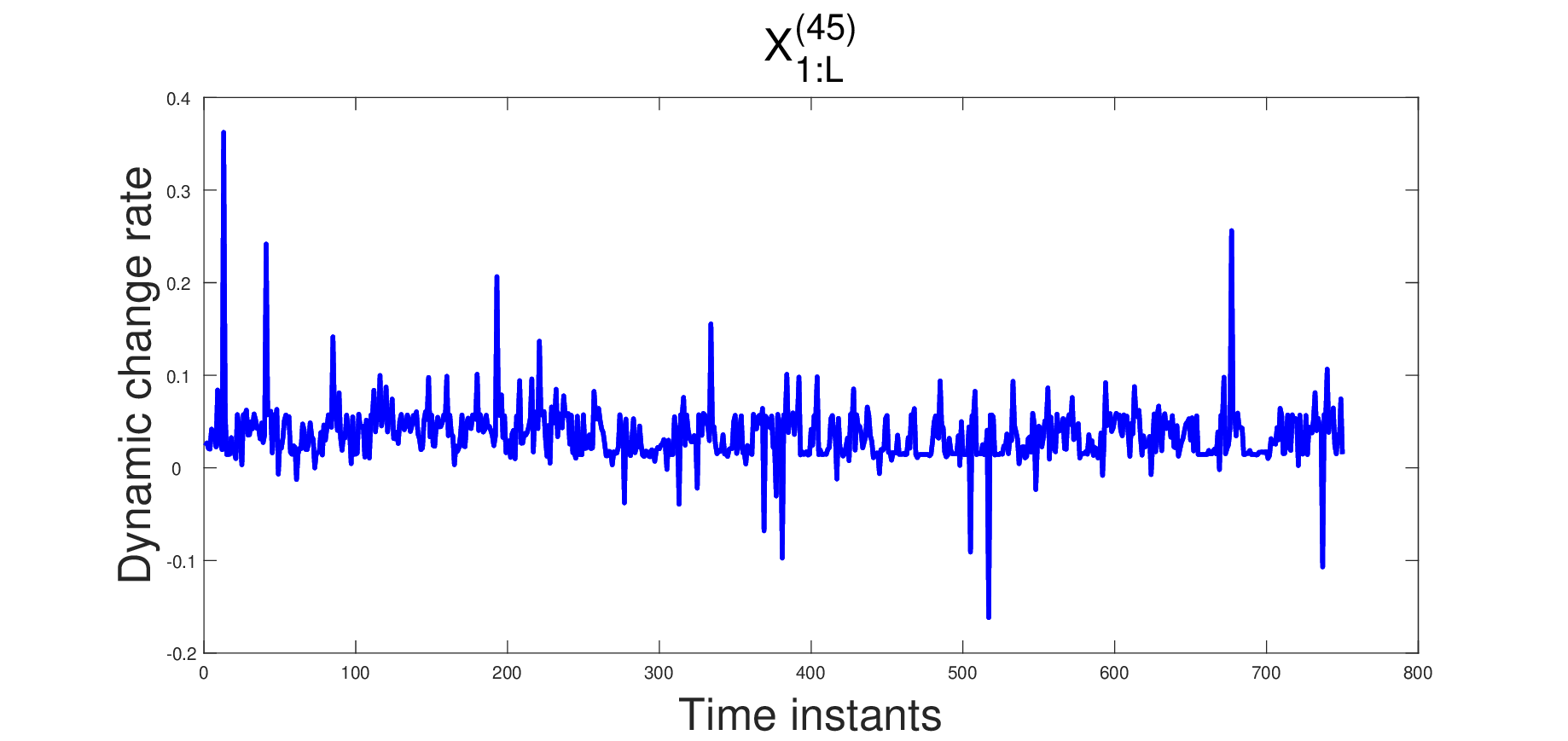}}
		\label{f1}
	\caption{Set of dynamic feature series after space smoothing and sequential general variational mode decomposition in a certain snapshot and one of the representative series.}
	\label{8}
\end{figure*}

Fig. \ref{8} shows the results of the dynamic feature extraction operation after spatial smoothing of the snapshot matrix $X$ of certain data with a time window of $L$. Where $L$ is 100 and the number of smooth steps is 800, naturally each dynamic feature sequence has 800 complex values. Figure $(a)$ shows the arrangement of 100 sets of trend change feature sequences obtained after spatial smoothing on the matrix. Figure $(b)$ shows the trend features of one set of time series, which is the real part of complex dynamic features. Such features can also be considered as the main component; Figure $(c)$ shows the arrangement of the imaginary parts of the dynamic feature complex values on the matrix after spatial smoothing, where the periodic change sequence is represented by the imaginary parts of the complex values representing frequency information, while Figure $(d)$ shows the feature changes of one set of frequency information.

\begin{figure}[H]
	\centering
	\subfigure[]{
		\includegraphics[scale=0.25]{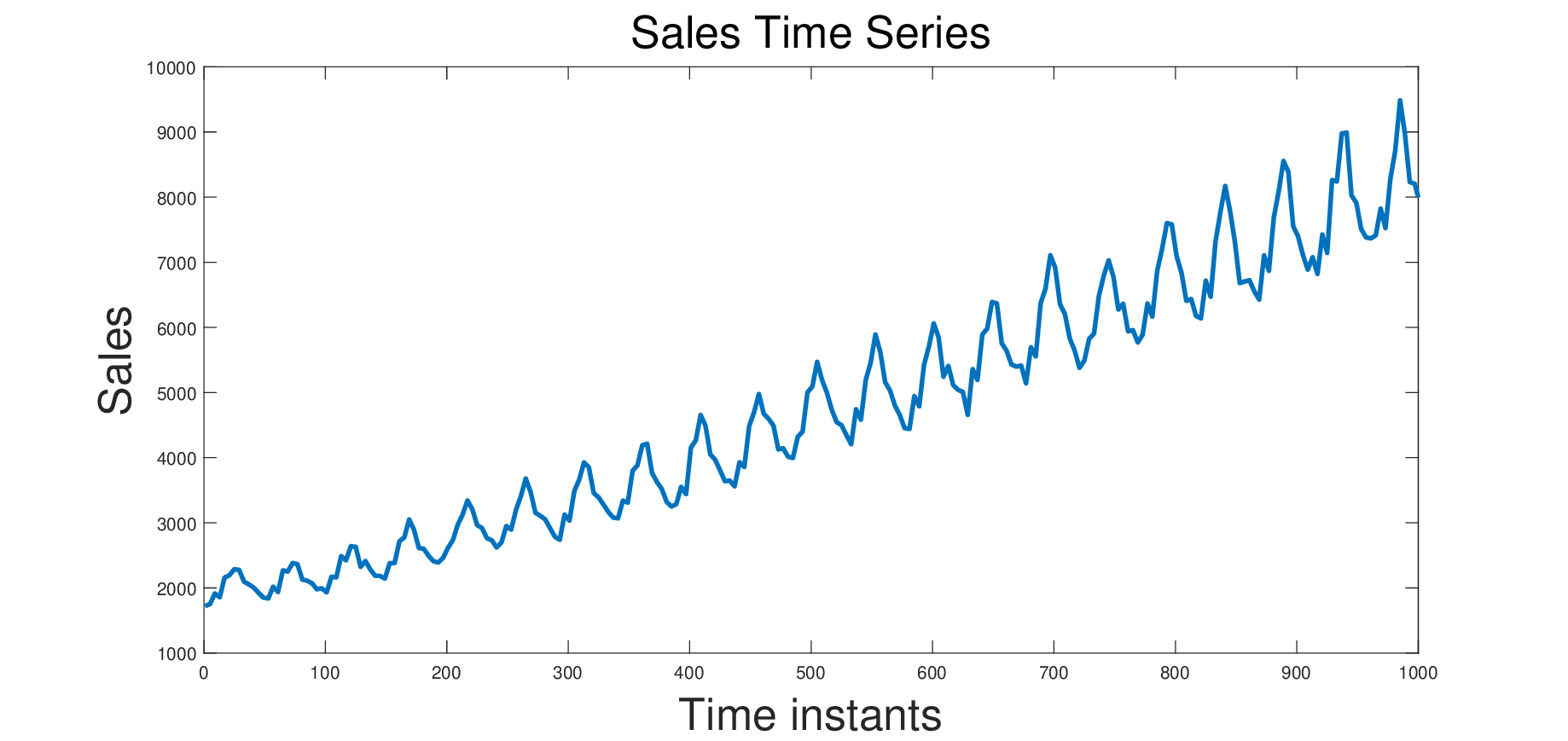}}
	\subfigure[]{
		\includegraphics[scale=0.25]{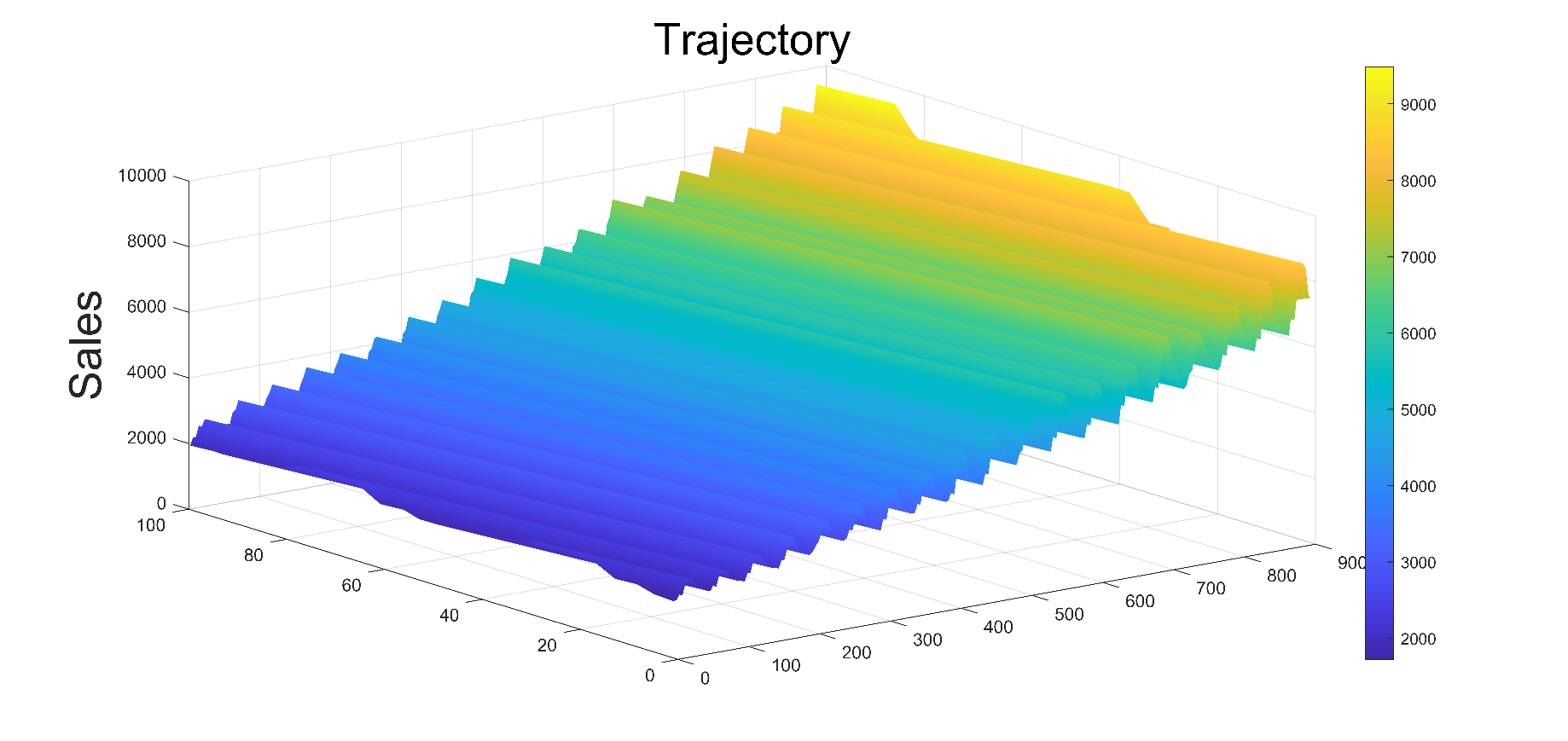}}
	\caption{Time series and its Trajectory Matrix of small amplitude oscillation trend seasonal periodic.}
	\label{7}
\end{figure}

After extracting a set of dynamic feature sequences for each snapshot, it is necessary to select and predict different parameters of the LSTM model for different types of dynamic feature sequences.

One of the main objectives of our research is to test and evaluate the proposed method using LSTM in this complex dynamic environment. We store 750 snapshots at equidistant time intervals. These snapshots are used to create a training foundation for our LSTM model, which is the training set of dynamic feature sequences used to train the LSTM network. To achieve this, we use a market sales time series of 1000 points as a window with a length of 100 to form the corresponding trajectory matrix. Fig. \ref{7} shows the seasonal sales sequence and trajectory matrix of a certain online store we used. This will result in 900 snapshots, of which the first 750 are used for LSTM model training to obtain a trained model. After being processed by the spatial smoothing algorithm, these snapshots will become a matrix with a column count of smoothing times and a row count of snapshot rows. These matrices will be divided into matrices dominated by trend terms, frequency terms, and residual terms. And what we need to do is predict the sequence of 750 points for each of these matrices and use the last 150 points as testing. In fact, there is no need to predict so many points, because after the prediction is completed, the matrix $V^*$ which is the predicted matrix can be transformed back into the predicted trajectory matrix, and the one-dimensional time series can be restored using the following formula:
\begin{equation}
	\begin{split}
		\hat{x}_i = 
		\left\{ 
		\begin{aligned}
		&\frac{1}{k}\sum_{p=1}^k V^*_{p,k-p+1}  &1\leq k \leq d^* \\
		&\frac{1}{d^*}\sum_{p=1}^{d^*}V^*_{p,k-p+1}  &d^*\leq k \leq m^* \\
		&\frac{1}{n-k+1}\sum_{p=k-m^*+1}^{n-m^*+1}V^*_{p,k-p+1}  &m^* \leq k \leq n
		\end{aligned}
		\right.
 	\end{split}
\end{equation}
The $V^*$ obtained from the LSTM neural network prediction is an $m \times d$ smooth result prediction matrix that needs to be further transformed into a time series. The diagonal averaging method, as a commonly used conversion algorithm, has accurate information conversion ability. Therefore, in this section, diagonal averaging is used to transform the predicted matrix $V^*$ to obtain a one-dimensional initial single component signal of length $n$. Finally, $d$ one-dimensional initial single component signals can be obtained, and the sum of all initial single components is the signal prediction result.

Assuming the element in $V_i$ is defined as $v_{ij}(1 \leq i \leq m, 1 \leq j \leq d)$. Set $d^* = min(m,d)$, $m^* = max(m,d)$ and $1 \leq i \leq m$, $1 \leq j \leq d$, if there is $m < d$, then $v^*_{ij} = v_{ij}$, otherwise $v^*_{ij} = v_{ji}$.

This article combines spatial smoothing algorithm and LSTM combination model to predict market sales. The spatial smoothing algorithm is used to decompose and calculate the snapshot sets under the track matrix formed by the dimension raising of the sales volume sequence to extract the dynamic feature series. LSTM combines the snapshot of the track matrix of the market sales volume and the dynamic feature sequence extracted from each line to conduct model training, and transforms the one-step prediction of the market sales volume into a unsupervised learning problem.

We will use the LSTM workflow described in Fig. \ref{4} to illustrate and handle different training paradigms. Implement and execute dynamic feature time series model prediction using LSTM network in MATLAB R2021b. In the experiment, the parameter settings of the LSTM network are shown in Table \uppercase\expandafter{\romannumeral1}.

\linespread{1.5}
\begin{table*}[b]
	\resizebox{\textwidth}{!}{
		\begin{tabular}{cccccc} \hline 
			\toprule
				Series Type	& Number of Hidden Layers & Initial Learn Rate & Maximum Epochs & Learn Rate Drop Period & Learn Rate Drop Factor\\
			\midrule
		Trend & 200 & 0.015 & 1000 & 350 & 0.01 \\  
			Frequency & 250 & 0.01 & 1500 & 300 & 0.015  \\  
		Residue & 300 & 0.005 & 2000 & 400 & 0.005\\
			\bottomrule
		\end{tabular}
	}
	\caption{LSTM network parameters for different dynamic feature types used for small amplitude oscillation time series.}
	\label{tab1}
\end{table*}

\begin{figure*}[htb]
	\centering
	\subfigure[]{
		\includegraphics[scale=0.25]{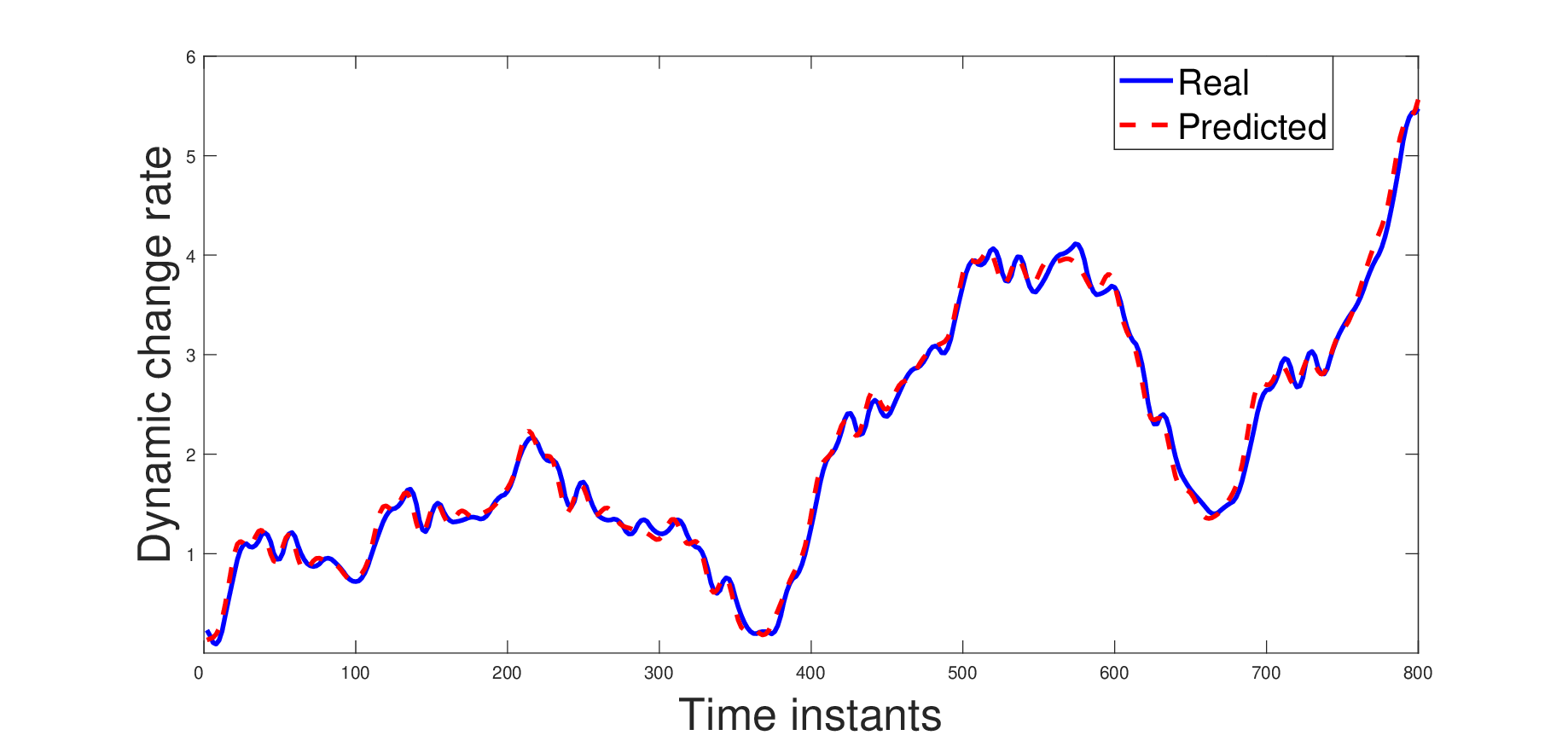}
		\includegraphics[scale=0.25]{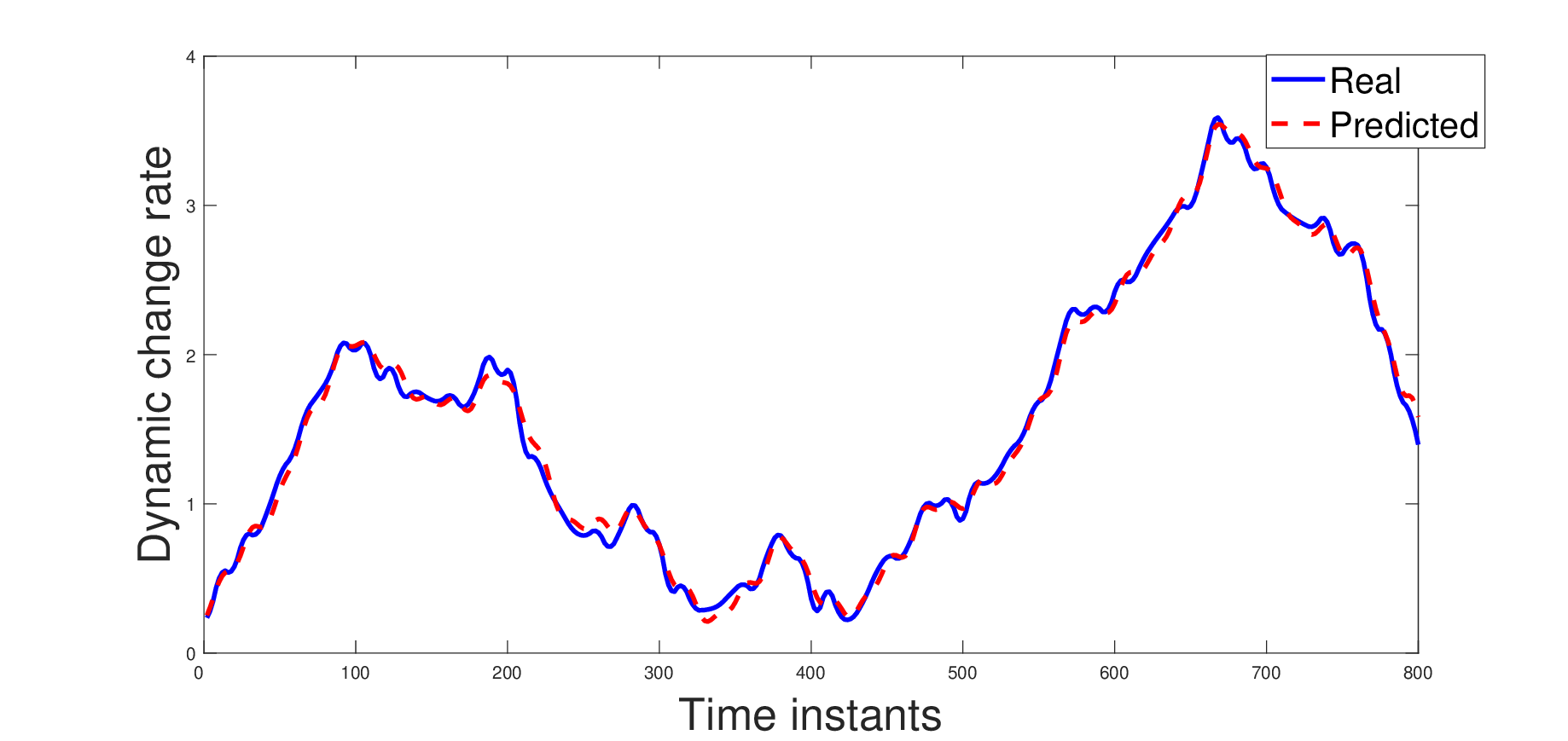}}
	\\
	\subfigure[]{
		\includegraphics[scale=0.25]{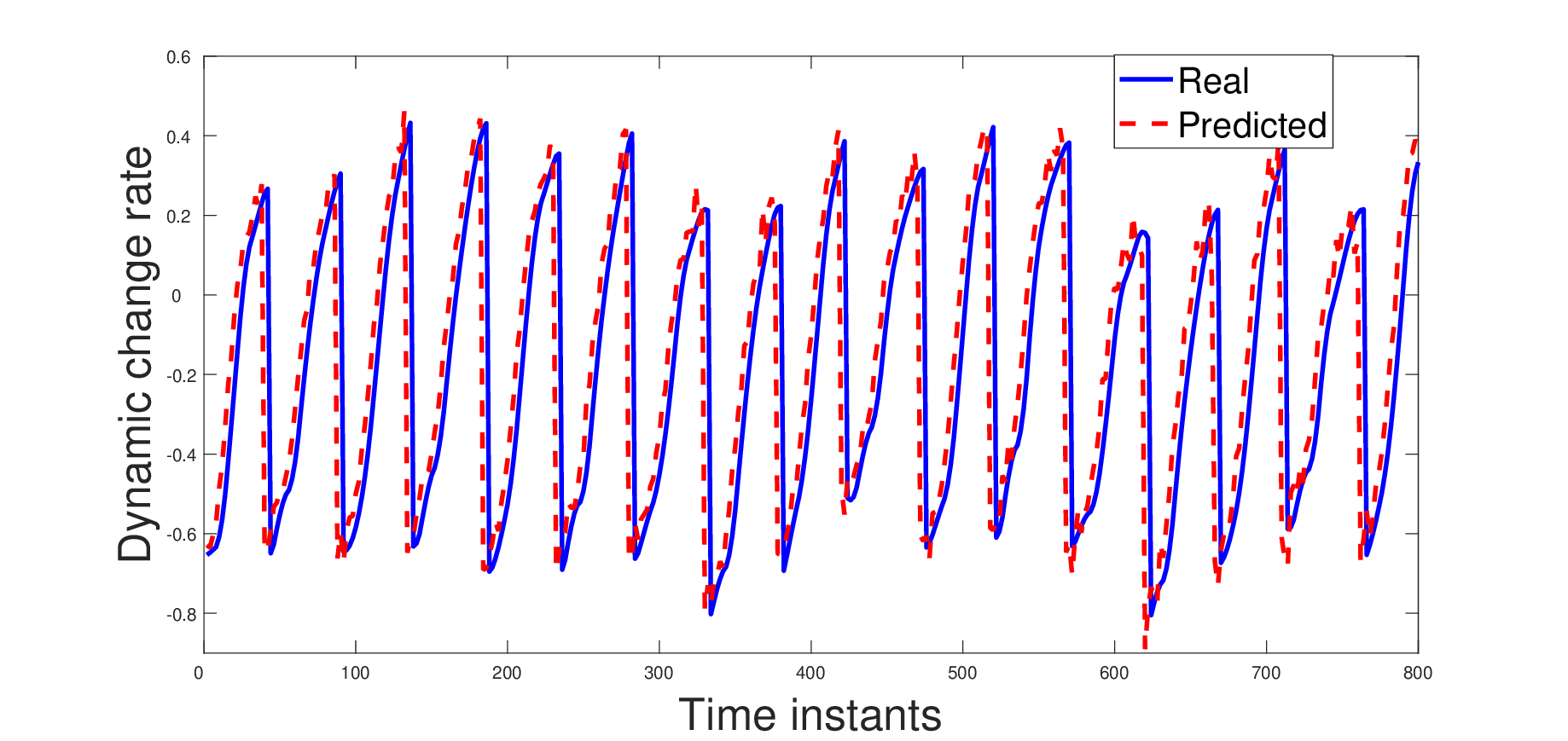}
		\includegraphics[scale=0.25]{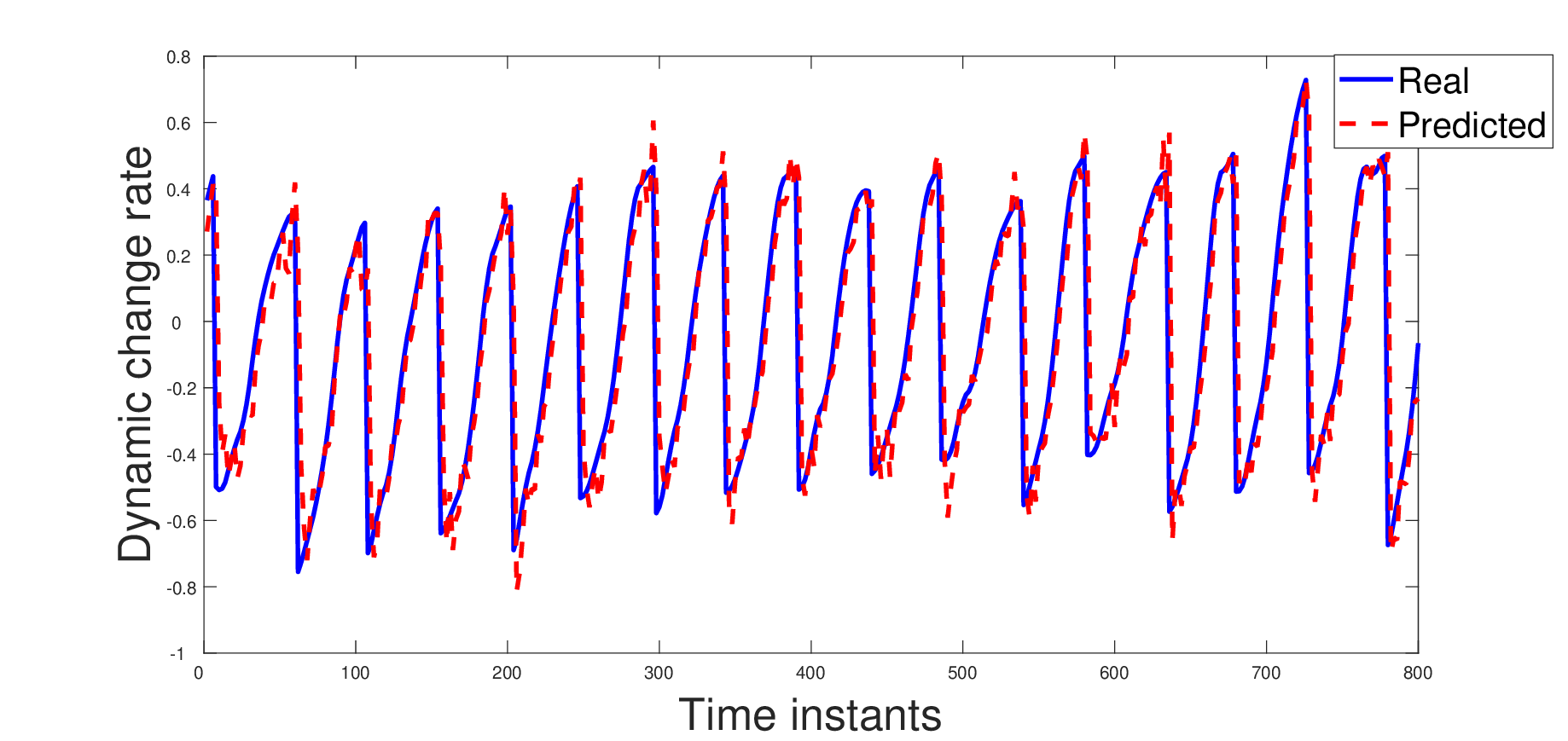}}
	\\
	\subfigure[]{
		\includegraphics[scale=0.25]{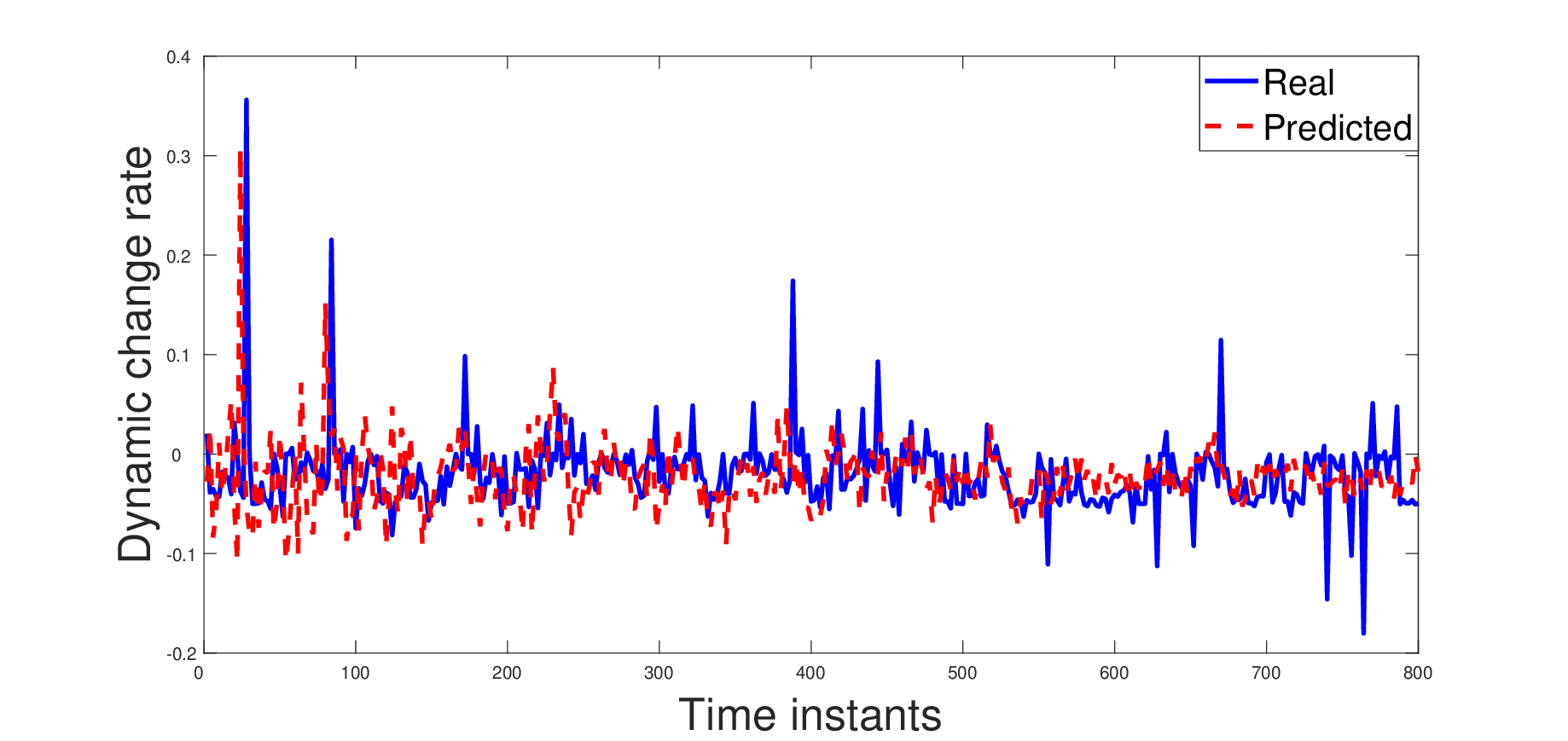}
		\includegraphics[scale=0.25]{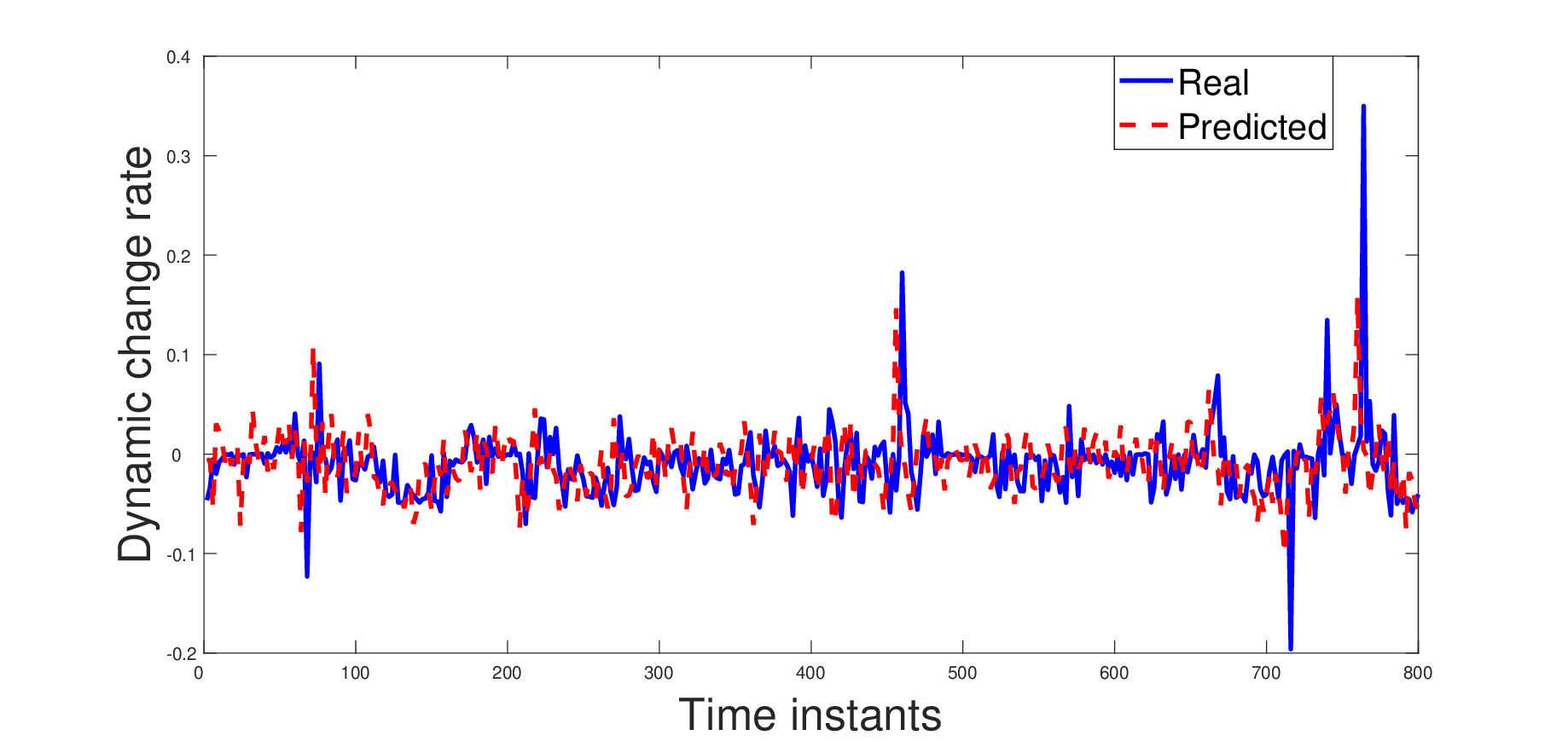}}
	\caption{Some representative trend, frequency, and residual dynamic feature series decomposed by sequential general variational mode decomposition and prediction results.}
	\label{9}
\end{figure*}

Apply the parameters proposed in Table 1 to the LSTM neural network to obtain prediction results on some typical dynamic feature change rate sequence samples. As shown in Fig. \ref{9}, it contains some typical trend features, frequency features, and residual prediction results.

\subsubsection{Large amplitude oscillation trend seasonal periodic time series}

The time series patterns of market commodity sales are mainly divided into unilateral patterns, oscillatory patterns, stationary patterns, and other specific patterns. The data changes vary under different market patterns, and the predictive performance of the model varies. The spatial smoothing algorithm mainly extracts modes based on historical trends of data, so there is a certain lag in the results obtained from modal decomposition. When the market is in a unilateral form, the spatial smooth extraction mode can accurately capture the trend information of market changes under the inertia of market trends; When the market is in a volatile market, the lag will weaken the tracking effect of dynamic features obtained from spatial smoothing on market trend changes. Therefore, this article will use a highly volatile dataset for empirical research on market patterns.

We selected the monthly beer sales volume of Steel Australia as the research object, and in order to meet the computational requirements of the spatial smoothing algorithm, we selected the parts with trend terms to construct a data trajectory matrix.

\begin{figure}[H]
	\centering
	\subfigure[]{
		\includegraphics[scale=0.25]{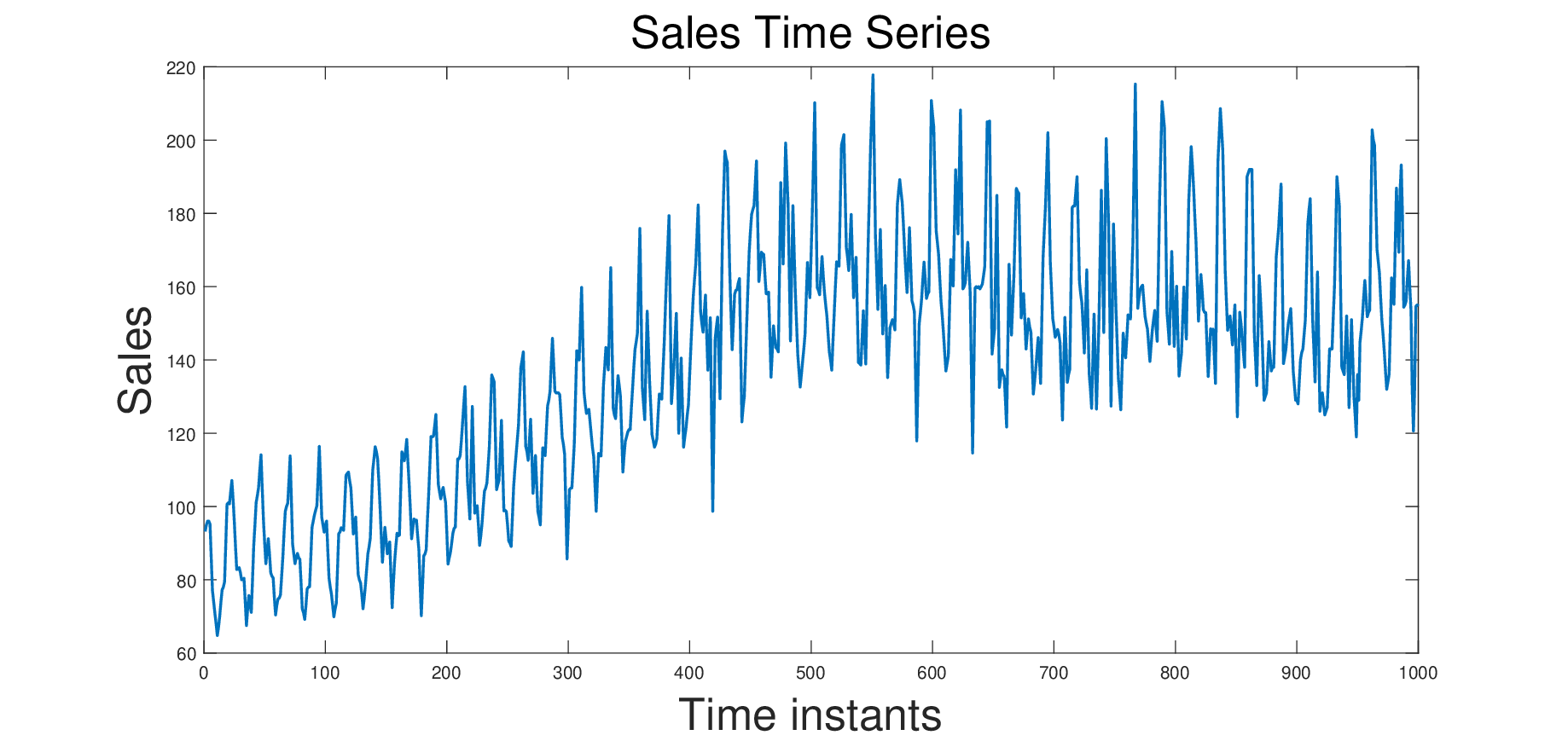}}
	\subfigure[]{
		\includegraphics[scale=0.25]{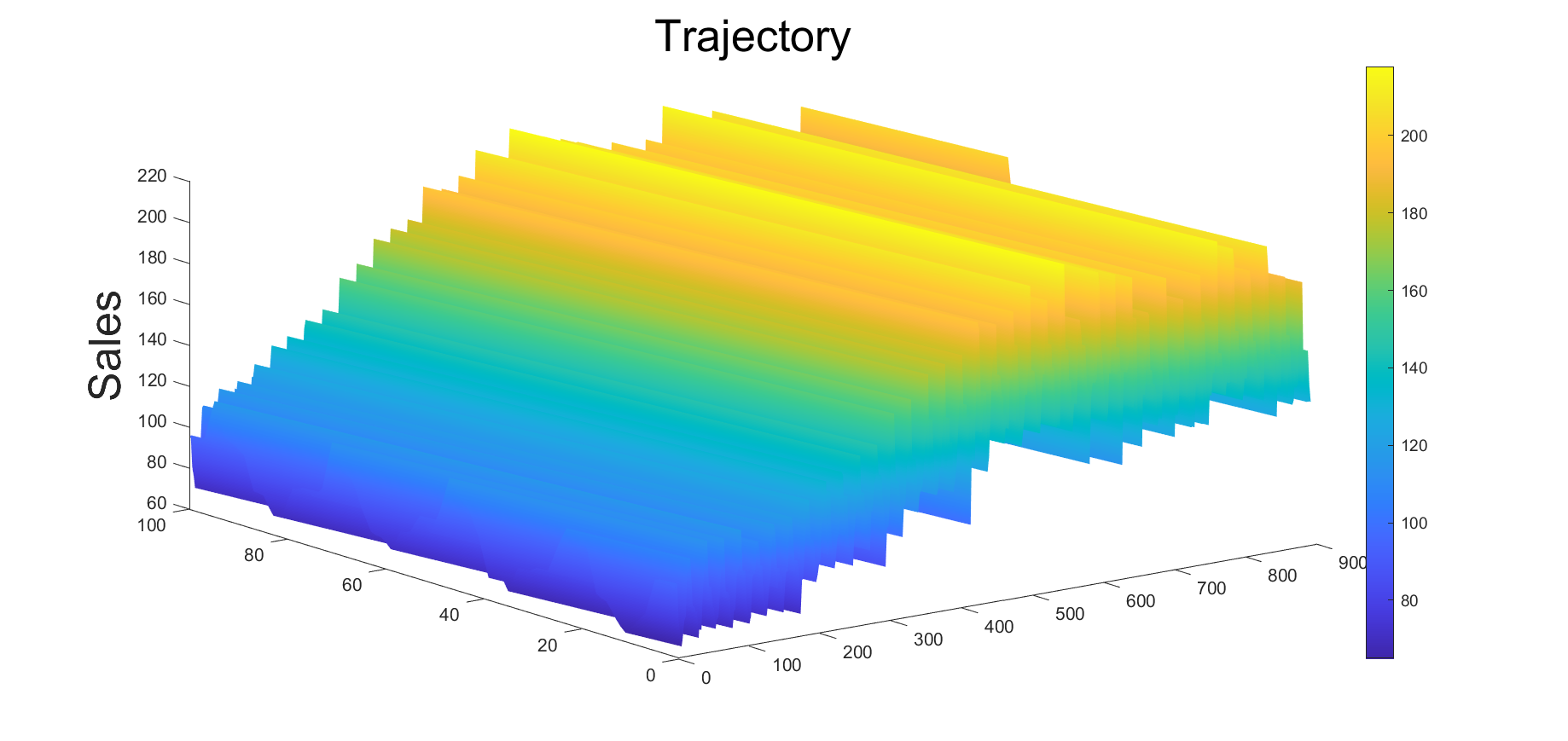}}
	\caption{Time series and its Trajectory Matrix of large amplitude oscillation trend seasonal periodic.}
	\label{10}
\end{figure}

It is not difficult to observe from Fig. \ref{10} that this time series not only exhibits periodic oscillations, but also exhibits less monotonic trend changes and complex seasonality. We believe that predicting such data will yield some unsatisfactory results.

\begin{figure*}[htb]
	\centering
	\subfigure[]{
		\includegraphics[scale=0.25]{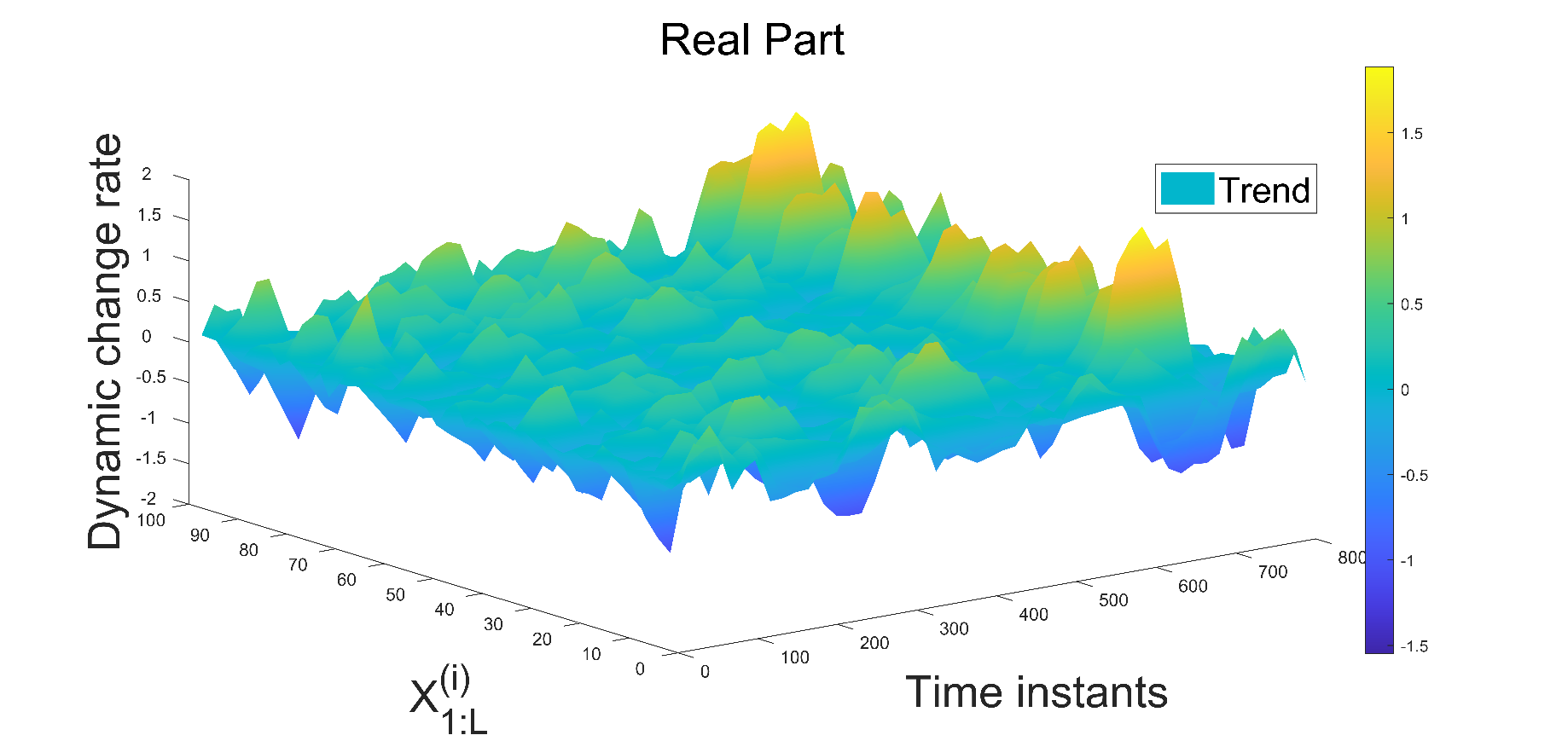}}
	\subfigure[]{
		\includegraphics[scale=0.25]{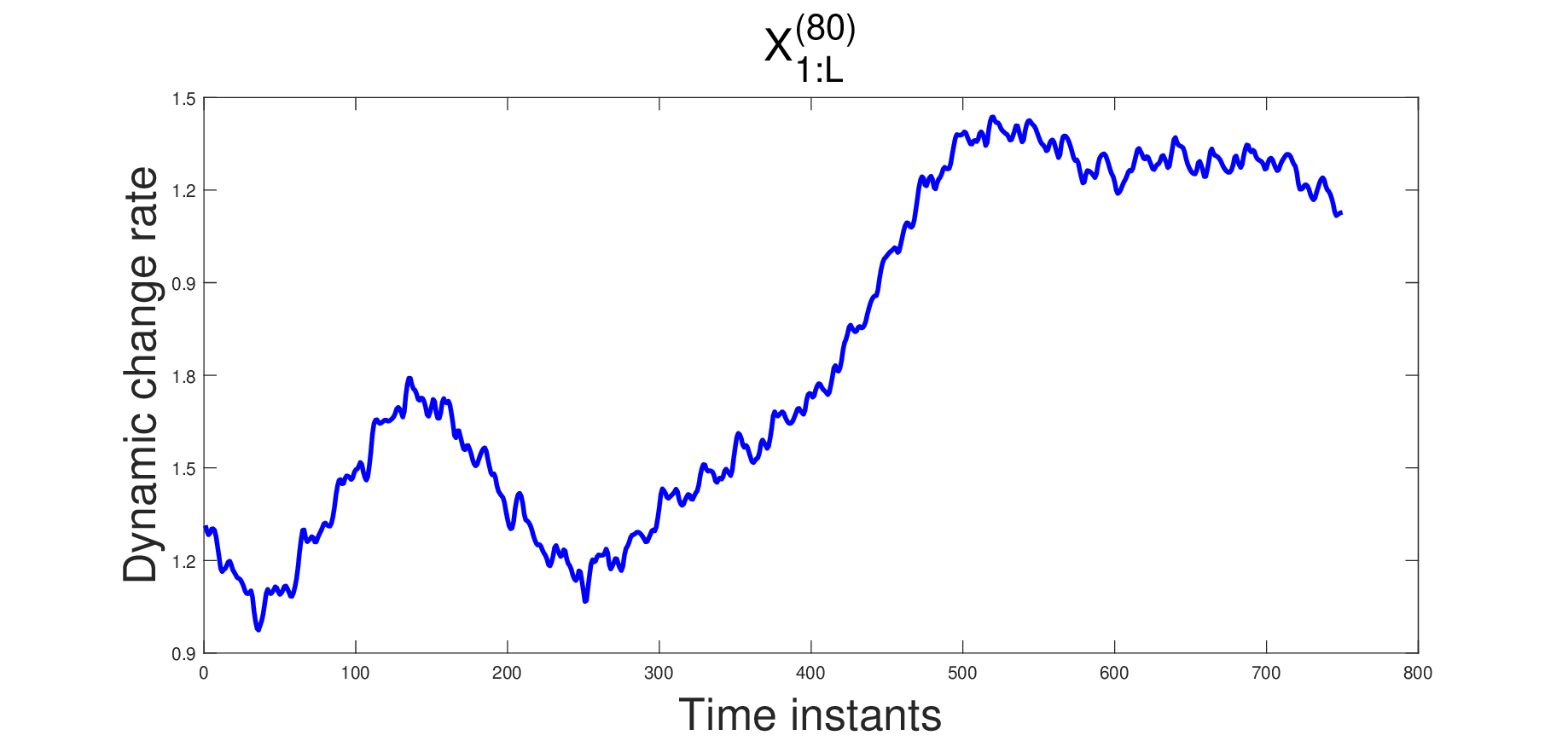}}
	\\
	\subfigure[]{
		\includegraphics[scale=0.25]{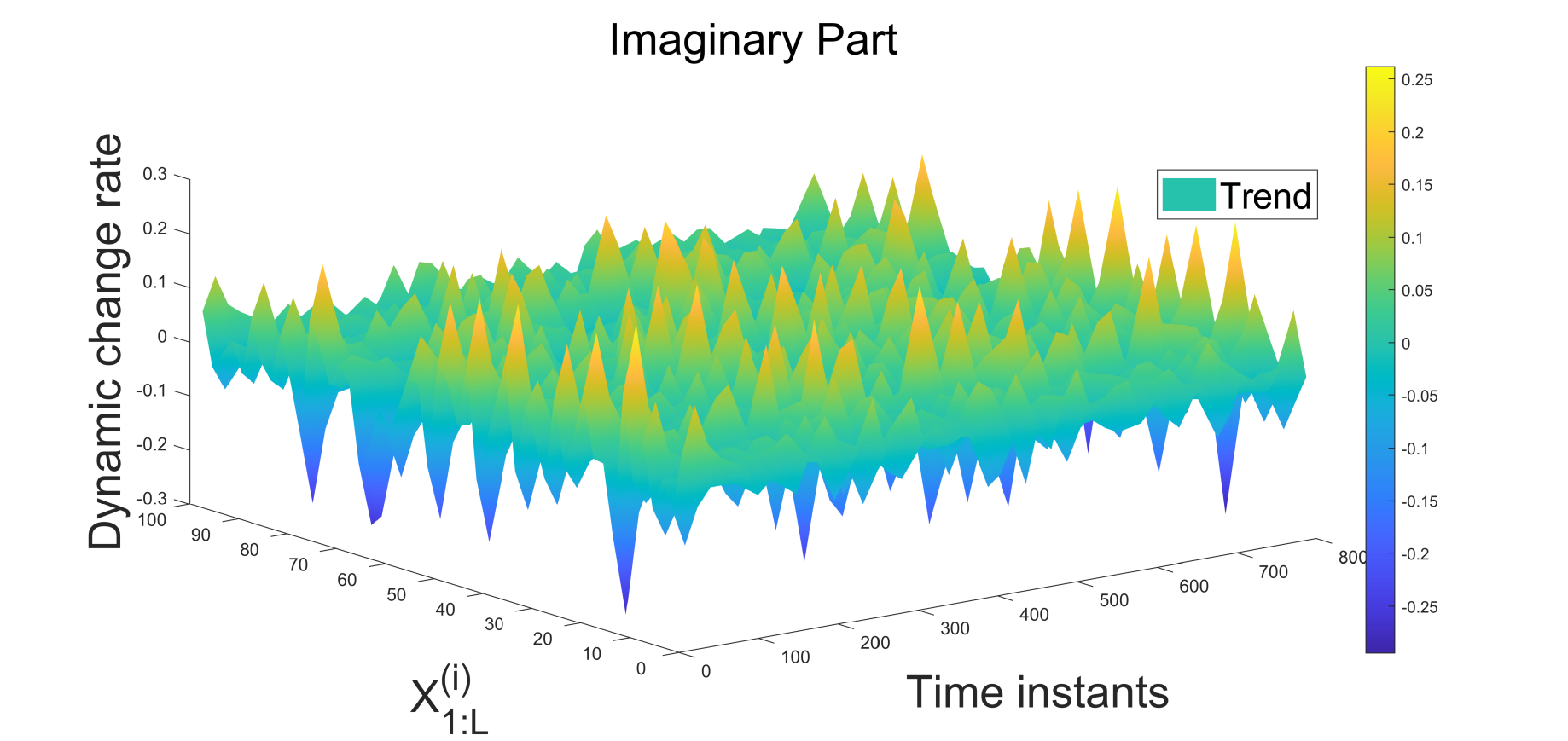}}
	\subfigure[]{
		\includegraphics[scale=0.25]{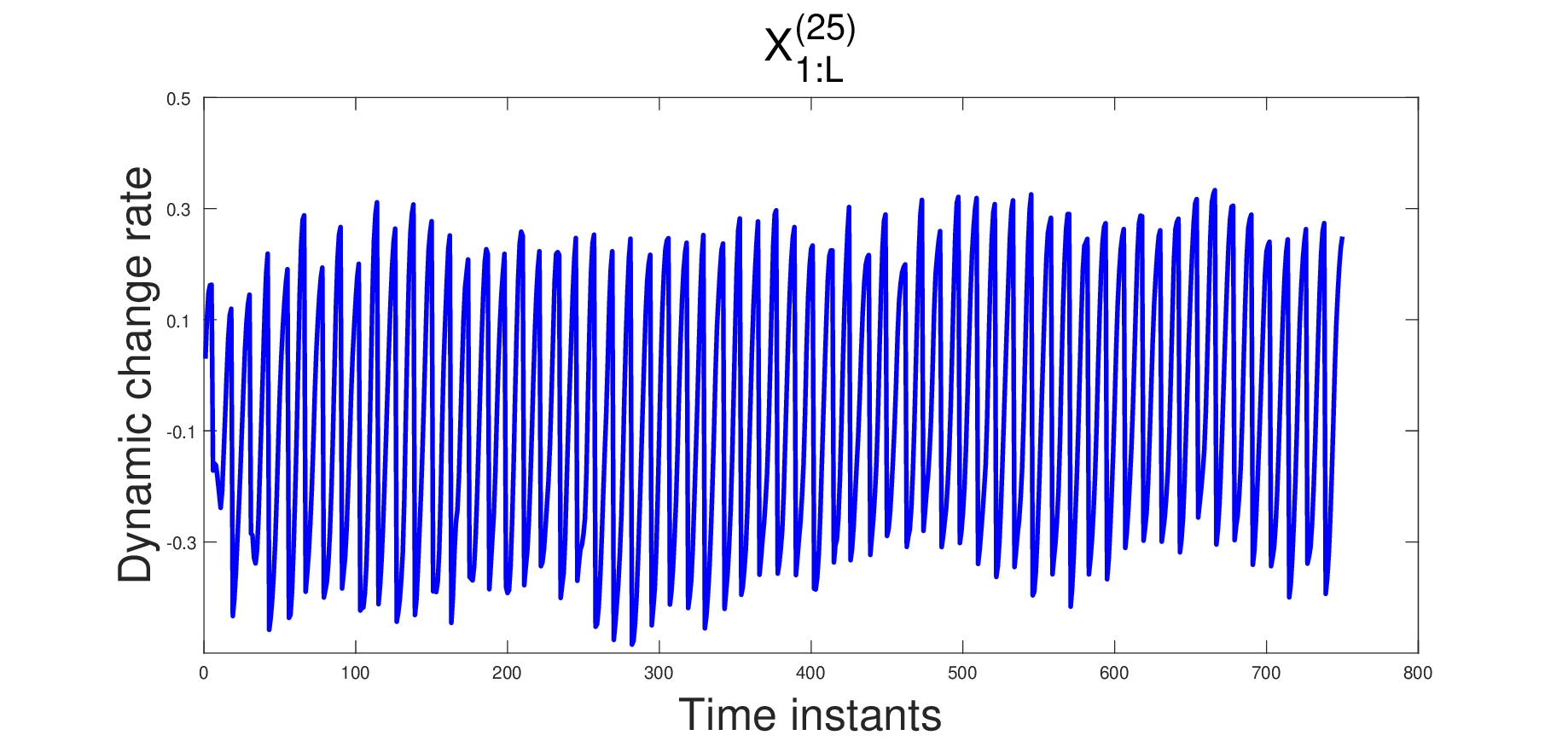}}
	\\
	\subfigure[]{
		\includegraphics[scale=0.25]{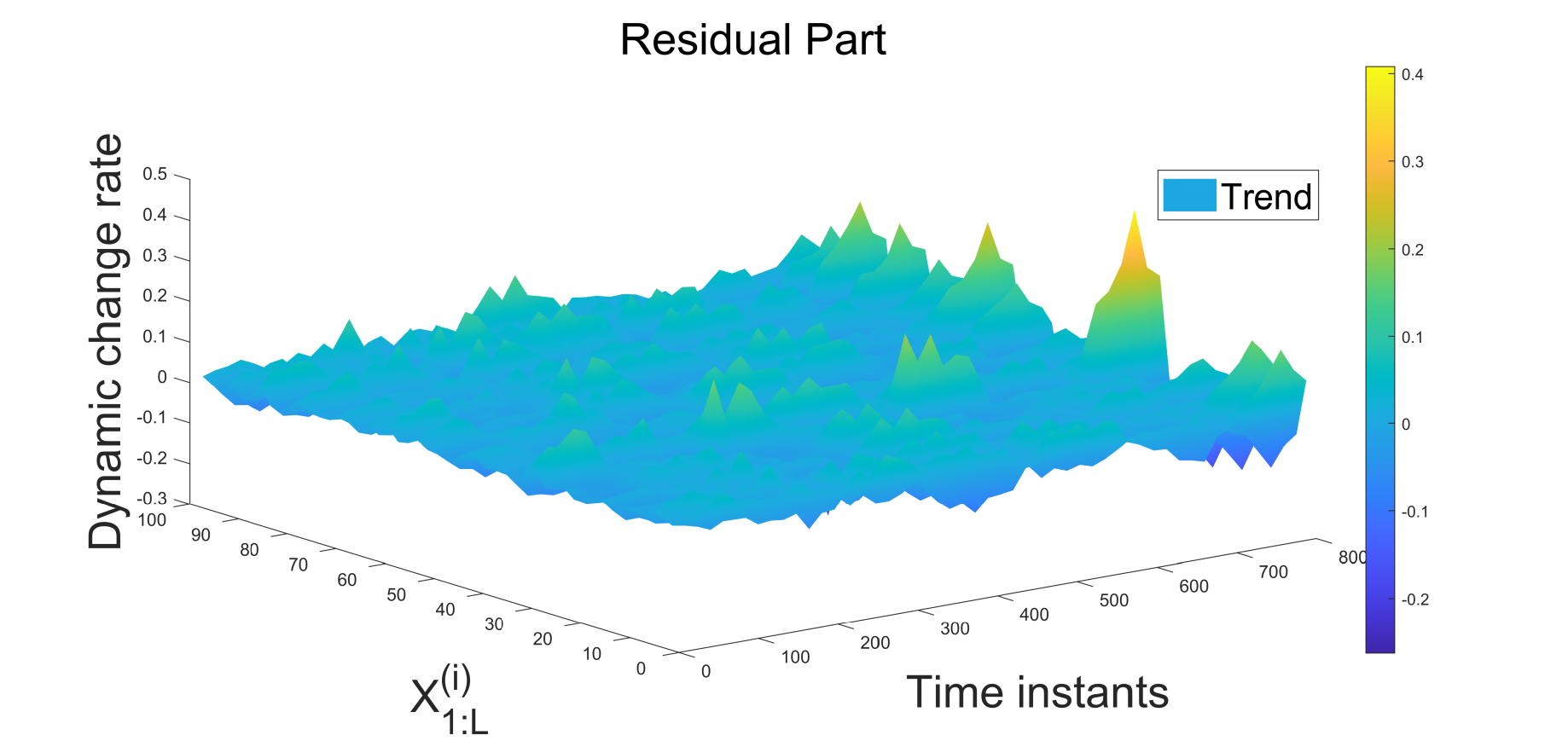}}
	\subfigure[]{
		\includegraphics[scale=0.25]{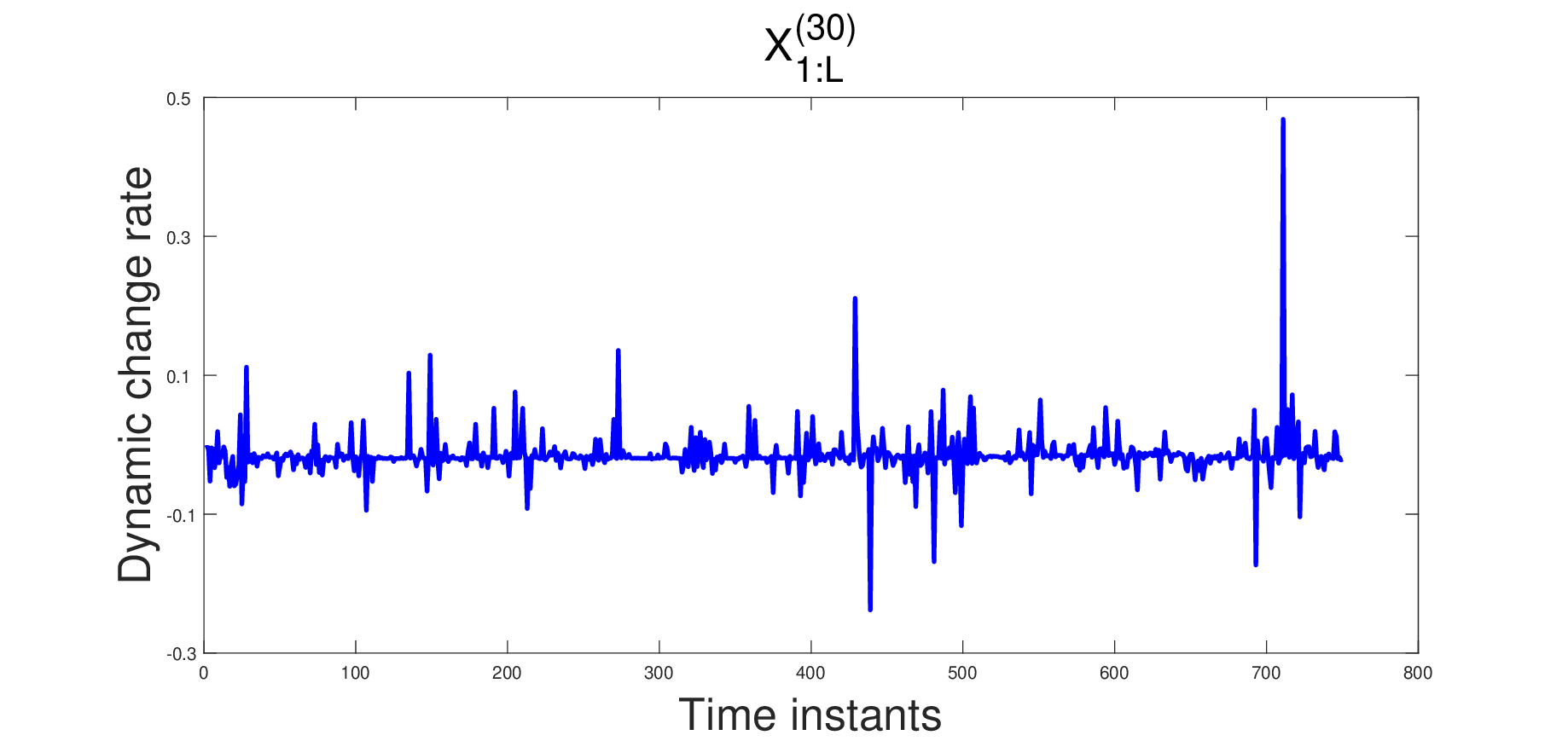}}
	\caption{Set of dynamic feature series after space smoothing in a certain snapshot decomposed by sequential general variational mode decomposition and some of the representative series.}
	\label{11}
\end{figure*}

After using spatial smoothing algorithm and feature extraction, it was found that the dynamic feature sequence with trend representation is a sequence with "rough edges" and some noise. The dynamic feature sequence representing the frequency part oscillates more violently. It is reflected in a larger amplitude of oscillation and a smaller period of oscillation, that is, a large amplitude and high frequency oscillation. However, with the help of neural networks, such oscillating data is generally easier to achieve good prediction results, as the errors transmitted back and forth that need to be identified and used are also periodic. For the residual part, like the previous data, it is quite troublesome because such data will have some large peaks, and for prediction algorithms, such peaks will generally be defaulted to outliers and ignored or added with very small weights. So when adjusting the prediction network parameters of such dynamic feature sequences, some changes and experience are needed.
\linespread{1.5}
\begin{table*}[t]
	\resizebox{\textwidth}{!}{
		\begin{tabular}{cccccc} \hline 
			\toprule
			Series Type	& Number of Hidden Layers & Initial Learn Rate & Maximum Epochs & Learn Rate Drop Period & Learn Rate Drop Factor\\
			\midrule
			Trend & 250 & 0.015 & 1200 & 350 & 0.01 \\  
			Frequency & 275 & 0.01 & 1500 & 325 & 0.015  \\  
			Residue & 300 & 0.005 & 1750 & 375 & 0.005\\
			\bottomrule
		\end{tabular}
	}
	\caption{LSTM network parameters for different dynamic feature types used for large amplitude oscillation time series.}
	\label{tab2}
\end{table*}

\begin{figure*}[htb]
	\centering
	\subfigure[]{
		\includegraphics[scale=0.25]{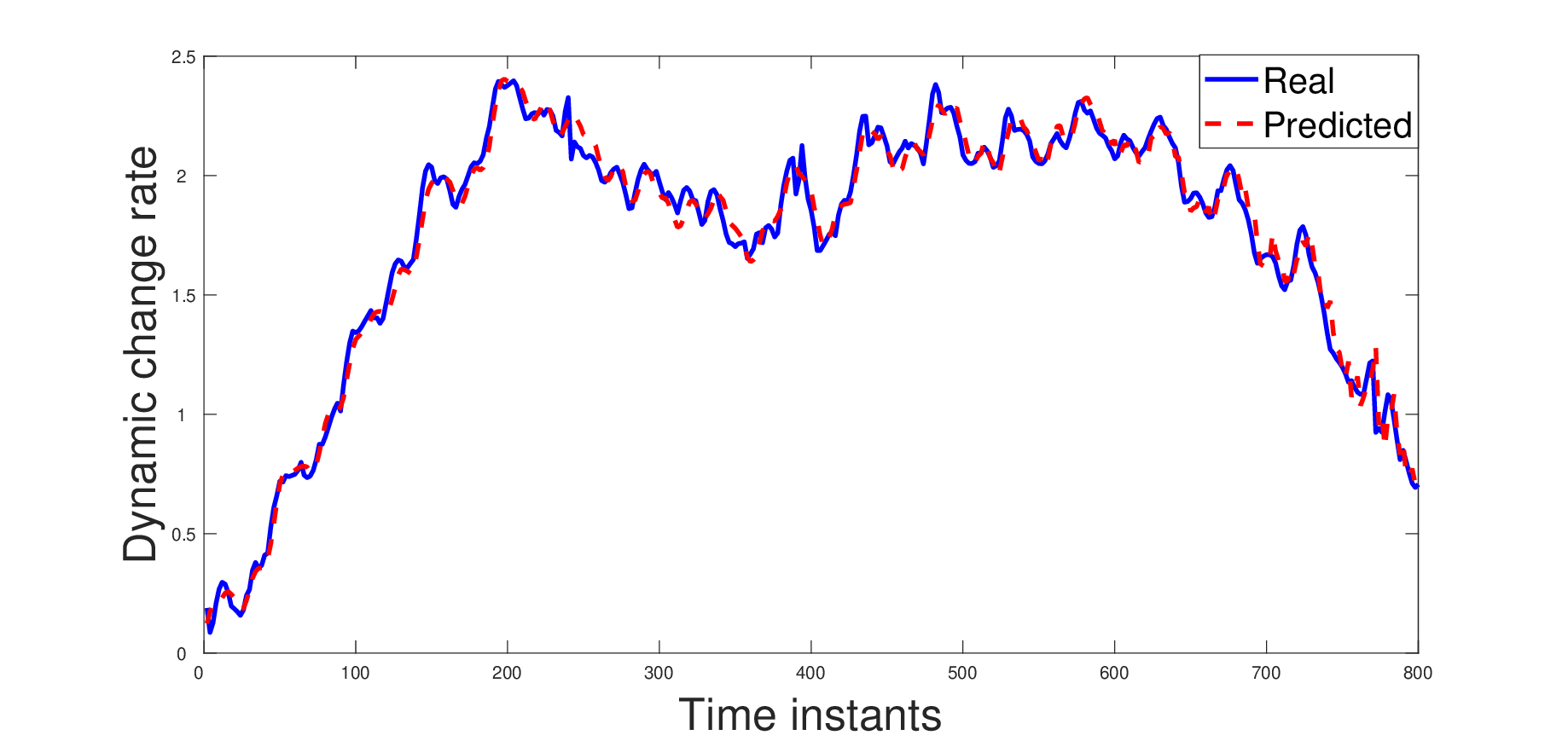}
		\includegraphics[scale=0.25]{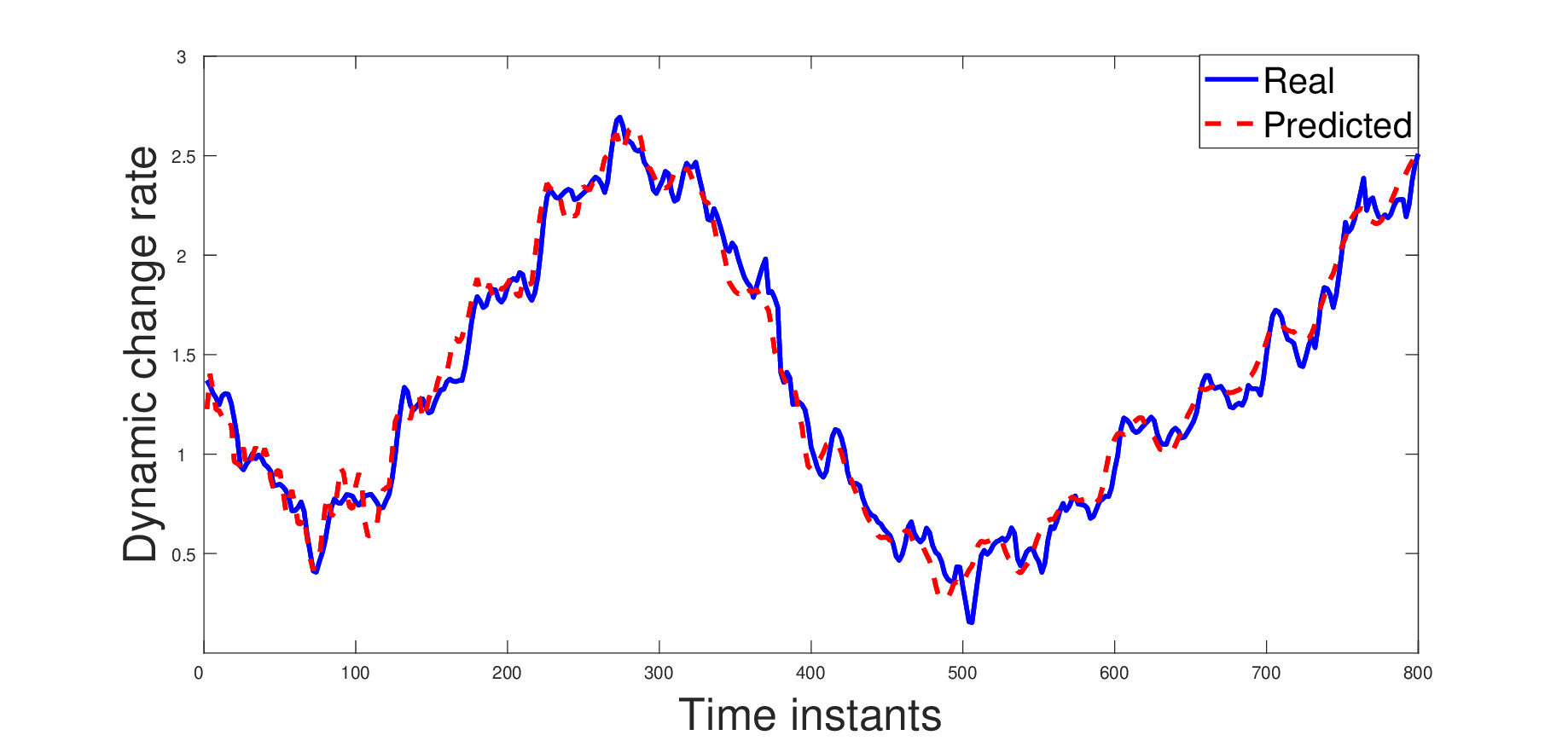}}
	\\
	\subfigure[]{
		\includegraphics[scale=0.25]{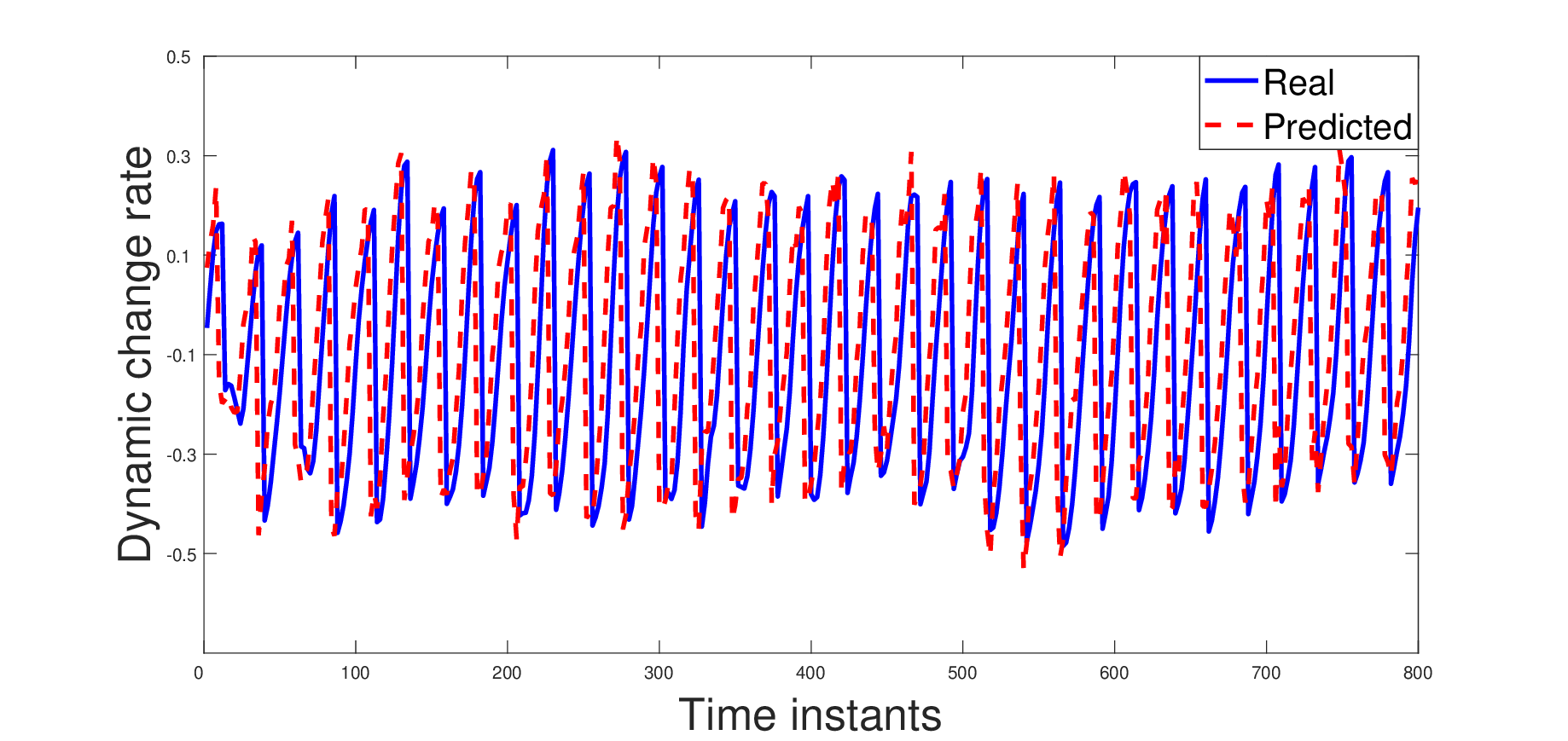}
		\includegraphics[scale=0.25]{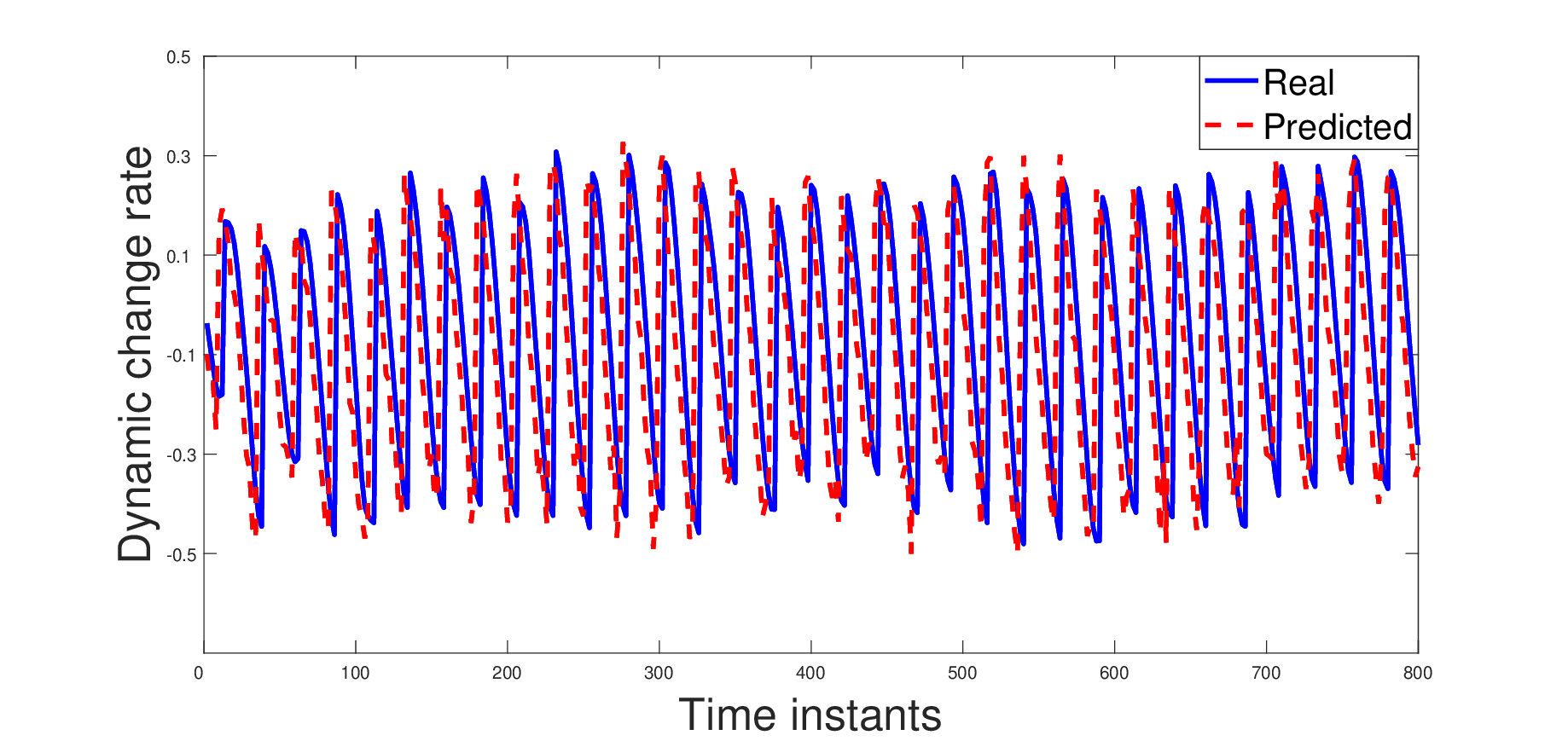}}
	\\
	\subfigure[]{
		\includegraphics[scale=0.25]{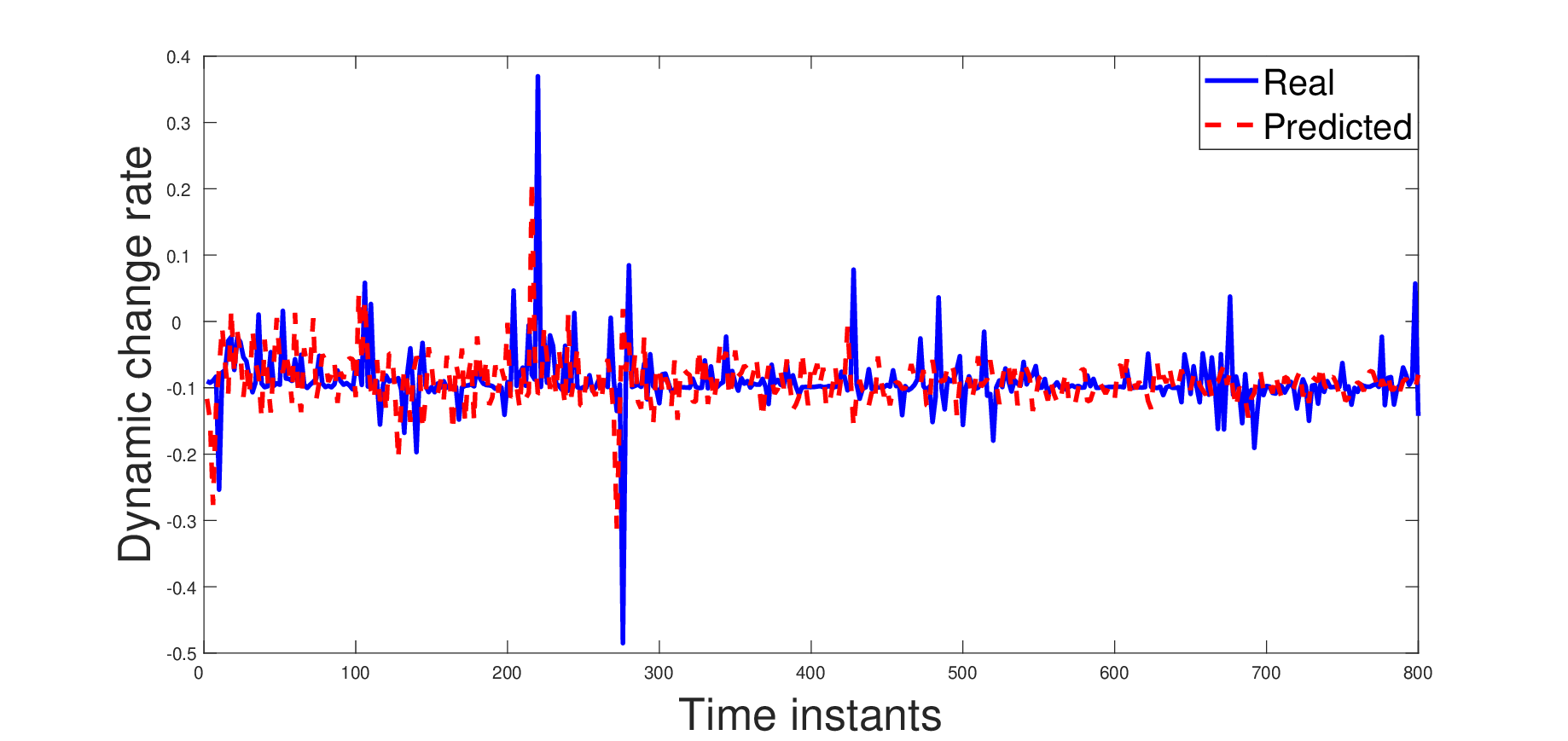}
		\includegraphics[scale=0.25]{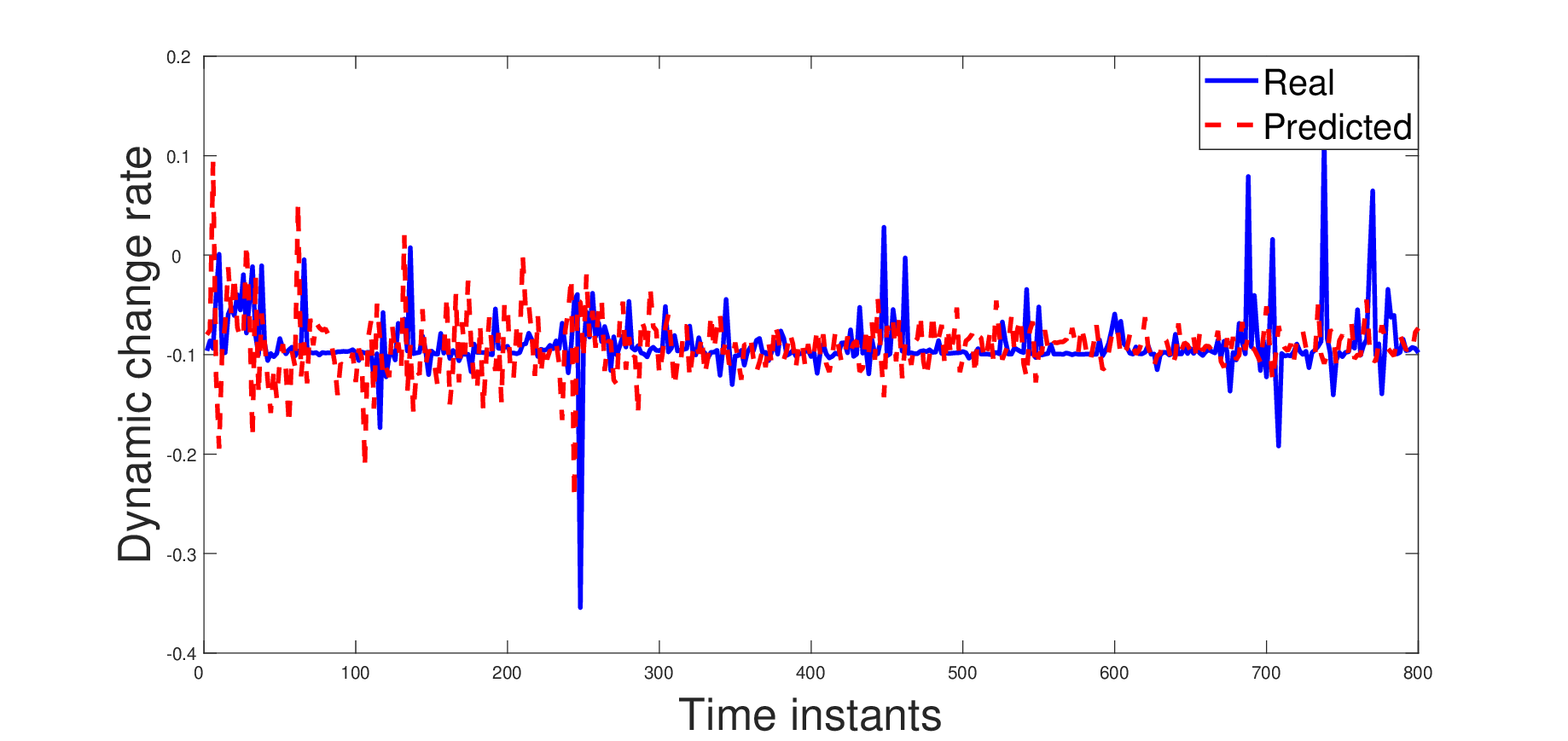}}
	\caption{Some representative trend, frequency, and residual dynamic feature series decomposed by sequential general variational mode decomposition and prediction results.}
	\label{12}
\end{figure*}

\subsection{Evaluation Indices}
When quantitatively evaluating the predictive performance of the model, we select the evaluation indicators for the sales forecasting ability as root mean square error (RMSE), mean absolute error (MAE), determine coefficient $R^2$ and mean absolute percentage error (MAPE). The smaller the RMSE, MAE, MAPE, the smaller the deviation of the model from the true value fitting, and the more accurate the results; The closer the coefficient R is to 1, the greater the goodness of fit and the better the model's ability to fit data.
\begin{equation} 
	\begin{split}
		RMSE\left( {y,\hat{y}} \right) = {\sqrt{\frac{1}{n}{\sum_{i = 1}^{n}\left( {y - {\hat{y}}_{i}} \right)^{2}}}}
	\end{split}
\end{equation}

\begin{equation} 
	\begin{split}
		MAPE\left( {y,\hat{y}} \right) = \frac{1}{n}{\sum_{i = 1}^{n}\left|\frac{y - {\hat{y}}_{i}}{y} \right|}
	\end{split}
\end{equation}

\begin{equation} 
	\begin{split}
		MAE\left( {y,\hat{y}} \right) = \frac{1}{n}{\sum_{i = 1}^{n}\left|{y - {\hat{y}}_{i}}\right|}
	\end{split}
\end{equation}

\begin{equation} 
	\begin{split}
		R^2\left( {y,\hat{y}} \right) 
		=\frac{SSR}{SST}
		=\frac{\sum\limits_{i = 1}^{n} \left( \hat{y}_i - \bar{y}\right) ^2}
		{\sum\limits_{i = 1}^{n} \left( y_i - \bar{y} \right) ^2}
	\end{split}
\end{equation}
Where $y$ is the actual value of the time series , $y_i$ is the real value, $\bar{y}$ is the mean of $y$, $\hat{y}$ is the predicted value, and $n$ is the number of data points in the time series.

\section{Results and Discussion}

In order to test the true predictive performance of the model and eliminate the influence of accidental factors, in the empirical study, 10 experiments were conducted on each model to compare the average results for such complex market sales situations.

The sales cycle fluctuates in a trend stage, with significant price fluctuations. The predicted results of the relevant models are shown in Table \uppercase\expandafter{\romannumeral3} and Fig. \ref{11}. Due to the severe lag in the prediction results, the combined model algorithm of decomposition and re prediction has significantly decreased the accuracy of price fitting prediction, with a negative goodness of fit, indicating that effective price prediction cannot be carried out. Compared to previous decomposition algorithms, the prediction accuracy of the DMD decomposition combination prediction model has been improved. However, due to the inaccurate information extracted from trend changes, the modal features are equivalent to introducing external noise and have not been smoothed, resulting in lower prediction accuracy than the proposed model. For such complex time series, pure use of LSTM neural networks would result in much better results, because the seasonal periodicity of such time series is predictable and easy to fit. The prediction effect of this example is superior to the spatial smoothing LSTM (SS-LSTM) model proposed in this article, as the extraction and prediction of dynamic features are more accurate in LSTM networks, resulting in good prediction performance for seasonal periodic time series with trends.
\begin{figure}[H]
	\centering
	\includegraphics[width=\linewidth]{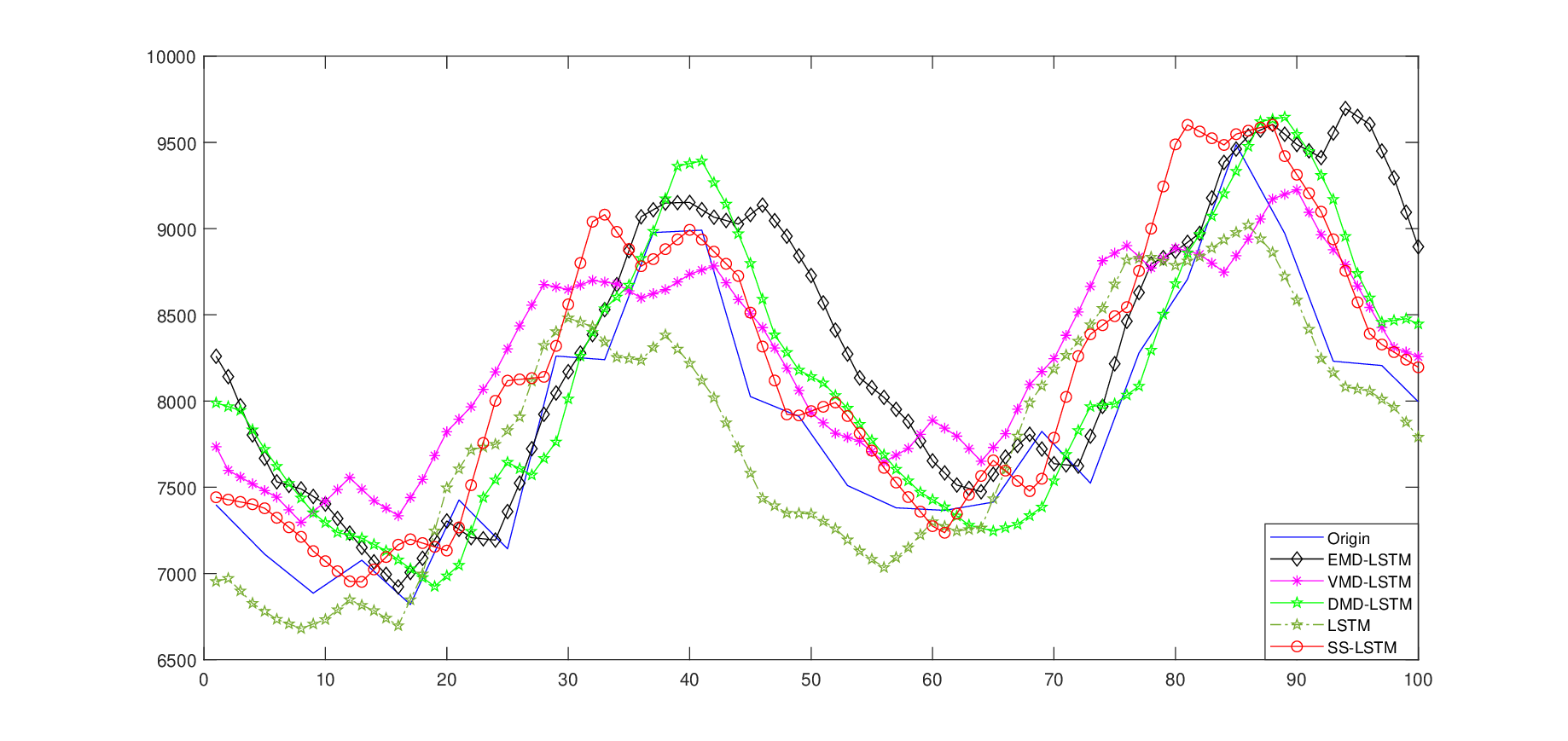}
	\caption{Comparison of prediction results of various models for small amplitude oscillation time series.}
	\label{13}
\end{figure}

After further averaging the predicted results for model fusion, it can be seen from the table that although the RMSE and $R^2$ of the predicted results have been slightly optimized, the impact of accidental factors cannot be ruled out. The empirical results indicate that the sales price prediction method based on the SS-LSTM model has advantages but still has certain limitations when applied in market cycles with trend fluctuations.

\linespread{1.5}
\begin{table}[H]
	\centering
	\begin{tabular}{ccccc} \hline 
		\toprule
		Model & RMSE & MAPE & MAE & $R^2$\\
		\midrule
		EMD-LSTM & 571.9884 & 0.0542 & 427.3596 & 1.7587 \\  
		VMD-LSTM & 484.3011 & 0.0549 & 424.3727 & 0.8839  \\  
		DMD-LSTM & 395.5711 & 0.0408 & 320.1567 & 1.3599 \\
		LSTM & 398.2447 & 0.0426 & 337.7391 & \textbf{1.0472} \\
		SS-LSTM & \textbf{390.0631} & \textbf{0.0388} & \textbf{308.4445} & 1.4695 \\
		\bottomrule
	\end{tabular}

	\caption{Comparison of specific values of evaluation indicators for prediction results of various models for small amplitude oscillation time series.}
	\label{tab3}
\end{table}

The prerequisite for effective feature extraction using the SS-LSTM method is the continuity of sales trends. When the market enters a period of fluctuation adjustment and the industry sector linkage effect is weak, it is difficult for the SS-LSTM algorithm to decompose effective trend information from short-term time series data.

\begin{figure}[H]
	\centering
	\includegraphics[width=\linewidth]{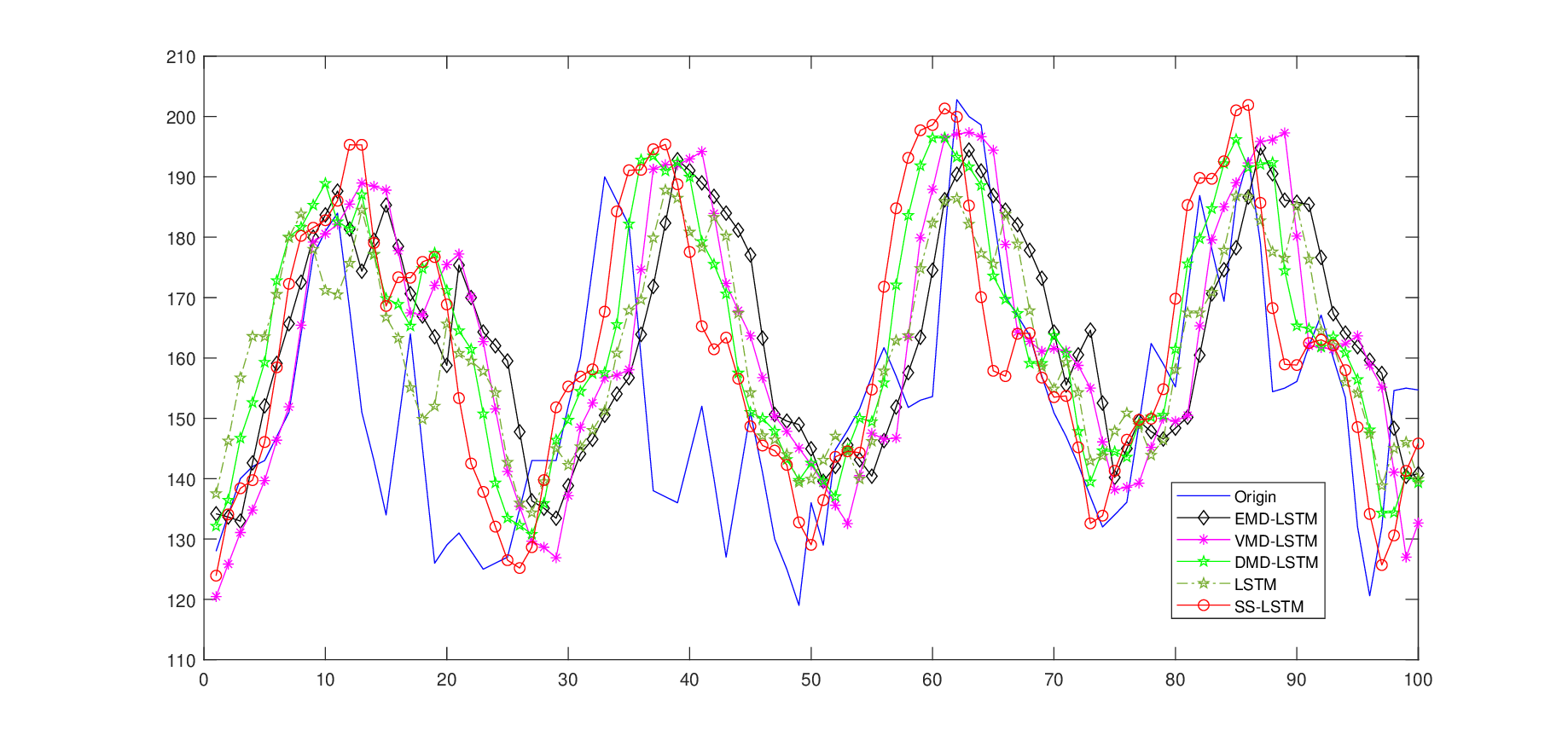}
	\caption{Comparison of prediction results of various models for large amplitude oscillation time series.}
	\label{14}
\end{figure}

Obviously, because the original data has some small peaks before and after each oscillation, the neural network will treat the center of gravity of a whole oscillation as a complete oscillation when predicting each component, making the $R^2$ value of this experiment not ideal. However, this is also a problem that every algorithm has. However, it can be observed that our proposed algorithm has the intention of increasing the weight of small peaks during the decay of the second peak, so that there is a local minimum value in its prediction process to fit such small peak oscillations. We believe this is a good adaptive phenomenon. However, from the table, our method is only slightly superior to other methods. So it is still difficult to predict time series with large amplitude and high frequency oscillations, especially for existing single time prediction algorithms. However, in the feature extraction part, we smoothed out the relatively independent parts through the relevant parts in the coherent signal, which is the advantage of our algorithm's robustness and universality. Compared to the features extracted by deep learning neural networks, such features are more interpretable, organized, and can confidently predict and recover signals.

The volatility effect is closely related to these evaluation criteria, as most evaluation criteria revolve around their stationarity with the center. Perhaps we can add some constraints and optimize them through some convex methods. For example, our corresponding author Chen proposed the idea of clustering constraints, which may enable better performance around the central frequency of the signal.

\linespread{1.5}
\begin{table}[H]
	\centering
	\begin{tabular}{ccccc} \hline 
		\toprule
		Model & RMSE & MAPE & MAE & $R^2$\\
		\midrule
		EMD-LSTM & 23.6071 & 0.1313 & 19.0307 & \textbf{1.1305} \\  
		VMD-LSTM & 24.1542 & 0.1296 & 18.7081 & 1.4013 \\  
		DMD-LSTM & 21.5996 & 0.1121 & 16.2862 & 1.2884 \\
		LSTM & 20.3701 & 0.1114 & 16.2749 & 0.8358 \\
		SS-LSTM & \textbf{20.2986} & \textbf{0.0990} & \textbf{14.6187} & 1.4506 \\
		\bottomrule
	\end{tabular}
	
	\caption{Comparison of specific values of evaluation indicators for prediction results of various models for large amplitude oscillation time series.}
\end{table}

\section{Conclusion and Prospect}

The market sales price prediction method based on the SS-LSTM model (Spatial Smoothing-Sequential General Variational Mode Decomposition-LSTM) proposed in this article starts from the complex non-stationary and trendy seasonal periodic time series, and uses Spatial Smoothing algorithm to extract dynamic features, and Sequential General Variational Mode Decomposition to obtain potential correlation factors and trend information brought about by seasonal periodicity. Then, the LSTM model is applied to specific numerical prediction of dynamic features, Fully utilizing temporal information and extracted dynamic features.

Through comparative experiments under the same background, it is shown that compared to traditional machine learning prediction methods, the model in this paper can achieve more accurate prediction results, including lower prediction error and higher directional accuracy, under the premise of significant market trends. This indicates that in market conditions with little change, seasonal correlation effects are more significant, and the use of spatial smoothing to extract dynamic features can effectively improve the predictive performance of the model. In a volatile market, the model may not perform well due to significant fluctuations in market sales, complex factors affecting price changes, weak linkage effects between industry sectors, and difficulty in obtaining effective trend information through relevant dynamic feature extraction methods.

For future prospects, we would prefer to address some aliasing components, whether in the spectrum or time-frequency domain, and we hope to separate them cleanly. This algorithm has a physical lower bound, but we can still use more elegant methods, such as projecting it onto a high-order space to separate it or projecting it onto other parallel spaces of the same dimension, which may require our team to study in a new signal space.

\end{multicols}

\end{document}